%% file: main.tex
\documentclass{article}
\usepackage{arxiv}
\usepackage{graphicx}
\usepackage[backend=biber, sorting=none, style=numeric, maxnames=1, url=false]{biblatex}
\usepackage{geometry}

\usepackage{amssymb}

\usepackage{tabularx}
\usepackage{array}
\usepackage{multicol}
\usepackage{authblk}
\usepackage{caption}
\usepackage{fancyhdr}
\usepackage{titlesec}
\usepackage{multicol}
\usepackage{breqn}
\usepackage{mathptmx}

\title{\emph{GATher}: Graph Attention Based Predictions of Gene-Disease Links}
\author[1,2]{David Narganes-Carlon\thanks{Corresponding author: \texttt{david.narganes95@gmail.com}}}
\author[1]{Anniek Myatt}
\author[1]{Mani Mudaliar}
\author[1,2]{Daniel J. Crowther}
\affil[1]{Exscientia, Oxford Science Park, OX4 4GE, UK}
\affil[2]{School of Medicine, University of Dundee, UK}

\date{}

\begin{document}

% Title
\maketitle

\input{00_abstract}
%TC:endignore

\newpage

\begin{twocolumn}
\section{Introduction}
\input{01_intro}
\section{Results}
\label{sec:results}
\input{02_workflow}
\input{03_eval}
\input{04_feats}
\input{05_explain}
\section{Discussion}
\label{sec:discussion}
\input{06_discussion}
\input{conclusion}

%TC:ignore
\section{Methods}
\label{sec:methods}
\input{07_data_nodes}
\input{07_data_edges}
\input{08_methods}
\input{acknowledge}

% Bibliography
\printbibliography

\section{Appendix}
\input{09_misc}

\end{twocolumn}
%TC:endignore

\end{document}

%% file: 00_abstract.tex
\begin{abstract}
Target selection is crucial in pharmaceutical drug discovery, directly influencing clinical trial success. Despite its importance, drug development remains resource-intensive, often taking over a decade with significant financial costs. High failure rates highlight the need for better early-stage target selection. We present \emph{GATher}, a graph attention network designed to predict therapeutic gene-disease links by integrating data from diverse biomedical sources into a graph with over 4.4 million edges. \emph{GATher} incorporates \emph{GATv3}, a novel graph attention convolution layer, and \emph{GATv3HeteroConv}, which aggregates transformations for each edge type, enhancing its ability to manage complex interactions within this extensive dataset. Utilising hard negative sampling and multi-task pre-training, \emph{GATher} addresses topological imbalances and improves specificity. Trained on data up to 2018 and evaluated through 2024, our results show \emph{GATher} predicts clinical trial outcomes with a ROC AUC of 0.69 for unmet efficacy failures and 0.79 for positive efficacy. Feature attribution methods, using Captum, highlight key nodes and relationships, enhancing model interpretability. By 2024, \emph{GATher} improved precision in prioritising the top 200 clinical trial targets to 14.1\%, an absolute increase of over 3.5\% compared to other methods. \emph{GATher} outperforms existing models like GAT, GATv2, and HGT in predicting clinical trial outcomes, demonstrating its potential in enhancing target validation and predicting clinical efficacy and safety.
\end{abstract}

%% file: 01_intro.tex
\subsection{Target Discovery}
Identifying and validating therapeutic targets is crucial in drug development due to high clinical failure rates and low drug approval rates \cite{Paul2010ImproveRD, Hutchinson2011, investment_2020, SerranoNjera2021TrendyGenesAC}. Developing a new drug typically takes over 10 years and requires substantial financial investment, with costs varying significantly depending on the therapeutic area, the complexity of the disease, and the success rates of early-stage trials \cite{Schuhmacher2023}. Effective target prioritisation is essential to improve these failure rates, focusing on factors such as safety, druggability, assayability, and novelty \cite{Emmerich2021GOTIT}. This strategy is vital for translating research from discovery to clinical trials and industry collaborations \cite{McDonagh2024HumanGA}. Enhancing early-stage target validation can significantly reduce phase II trial attrition by 24\% and lower development costs by 30\% \cite{Paul2010ImproveRD}.

Recent advances in artificial intelligence (AI) have shown great potential in expediting target identification and validation, thereby accelerating various aspects of the drug discovery process \cite{KPJayatunga2024HowSA}. Moreover, causal links derived from human genetics can help identify potential targets, facilitating the transition from laboratory research to clinical trials \cite{McDonagh2024HumanGA}. By 2023, over 30\% of the clinical pipeline included AI-discovered targets \cite{KPJayatunga2024HowSA}, highlighting the impact of computational tools. However, the novelty and efficacy of many of these targets remain under debate. Traditional resources such as Open Targets \cite{opentargets2023} primarily aggregate existing data, underscoring the need for computational tools that can predict and infer new relationships, thereby filling gaps and completing missing information in biomedical datasets \cite{NarganesCarln2023APA}.

\subsection{Biomedical Graphs}
The rapid growth of biomedical data and ontologies has transformed the construction of biomedical graphs, enabling the integration of diverse datasets into unified models \cite{Himmelstein2017, FernandezTorras2022, Chandak2023}. These graphs amalgamate data from genomics, transcriptomics, chemical screenings, electronic health records, and clinical trials. Given that biomedical data is inherently sparse and disconnected, there is a pressing need for integration and computational tools to prioritise and rank the most promising targets. Integrative graphs provide a comprehensive view by linking genes, pathways, phenotypes, drugs, and clinical trials, representing diseases as interconnected systems. These graphs are essential for computational methods, as they improve target identification and enhance our understanding of complex biomedical systems \cite{Himmelstein2017, FernandezTorras2022, Chandak2023}.

\subsection{Previous Work}
Graph-based approaches have demonstrated significant promise in biomedical research. For instance, Rosalind \cite{Paliwal2020} utilises tensor factorisation to rank genes across diseases and identify therapeutic disease-gene relationships, showing a performance improvement of 18\%-50\% over comparable algorithms. Similarly, Ye et al. \cite{Ye2021AKG} developed a tensor factorisation model with knowledge graph embeddings, significantly improving prediction accuracy. In contrast, Bioteque \cite{Himmelstein2017} employs node2vec methods for graph embeddings but does not implicitly learn relationships. However, none of these models couple their predictions with explainability algorithms, which are crucial for understanding and interpreting the relationships. The R2E model by BenevolentAI \cite{Patel2024RetrieveTE} predicts clinical trial outcomes using combined genetics and literature data from 2020, achieving ROC AUC (Receiver Operating Characteristic Area Under the Curve) of 0.545 for genetics-only data, 0.579 for literature-only data, and up to 0.638 for combined modalities. The ROC AUC metric is a measure of the model’s ability to distinguish between classes, where a higher value indicates better performance. These results highlight the importance of integrating diverse data sources for improved predictive performance.

\subsection{Graph Attention Networks}
Graph Attention Networks (GAT) introduced multi-head self-attention mechanisms, addressing limitations of spectral-based graph neural networks \cite{GAT2017}. GAT uses multi-head attention to enhance learning stability and performance. However, GAT applies the same transformation across all node types, potentially missing unique node characteristics. Graph Attention Network version 2 (GATv2) \cite{GATv22022} improves upon GAT by introducing separate projection matrices for source and target nodes, allowing dynamic attention. Heterogeneous Graph Transformer (HGT) \cite{HGT2020} further improves by using separate projection matrices for each node type and specific transformations for each edge type, handling different relationships within the graph.

\subsection{GATher, GATv3, and GATv3HeteroConv}
GATher is an end-to-end framework for target prioritisation, encompassing data integration into a biomedical knowledge graph, a new attention layer per edge type (Graph Attention Network version 3 (GATv3)), an aggregation layer with cross-relationship attention (GATv3HeteroConv), a combination of engineered and learnable node features, pre-training/fine-tuning, and custom negative sampling.

The biomedical graph contains over 4,456,990 directed edges from sources like MONDO \cite{Vasilevsky2022MondoUD}, Human Phenotype Ontology (HPO) \cite{Gargano2023TheHP}, Gene Ontology (GO) \cite{Aleksander2023TheGO}, Reactome \cite{reactome2021}, Kyoto Encyclopedia of Genes and Genomes (KEGG) \cite{Kanehisa2023}, Pharmaprojects \cite{Pharmaprojects2023}, and Open Targets \cite{opentargets2023}. The attention mechanism in GATv3 considers shared interactors among biological nodes, identifying nodes similar if linked to common diseases, functions, or interactors \cite{Kovacs2019Network, Eyuboglu2022MutualIP}.

GATher uses a GATv3 layer for each edge type, combined with GATv3HeteroConv to integrate diverse relationships, enhancing interpretability and performance. An attention mechanism across relationships weights the importance of each edge type, providing insights into the significance of different relationships. The training schedule involves pre-training to predict edge types and fine-tuning on clinical trial outcomes to calibrate efficacy predictions. In the decoder, the model regresses the maximum clinical trial phase for each outcome type, using estimates of safety and positive efficacy to prioritise targets. The model can also be employed for disease recommendation when seeking indication expansion.

Evaluation used a retrospective 2018 test set and a prospective 2024 test set. Hard negative sampling mitigates over-recommendation of popular targets and diseases \cite{bonner2022implications}. GATher integrates engineered features like gene expression profiles, molecular fingerprints, and learnable embeddings capturing topological patterns. Features from the Human Protein Atlas \cite{Sjstedt2020AnAO} and Decipher \cite{Collins2022, Karczewski2020} provide spatial, abundance, and essentiality information.

Graph explainability algorithms applied to GATher provide visual explanations by highlighting influential nodes and edges, validating predictions and revealing critical subgraphs.

Integrating trial data on outcomes like unmet efficacy, positive efficacy, and safety-related terminations, GATher improved therapeutic target prioritisation in test sets from 2018 and 2024. GATher achieved a ROC AUC of 0.69 for unmet efficacy failures and 0.79 for positive efficacy outcomes for any phase transition. Specifically, for Phase 2 to Phase 3 transitions, GATher achieved a ROC AUC of 0.67 for unmet efficacy failures and 0.81 for positive efficacy outcomes, outperforming HGT, GAT, GATv2, and the R2E model \cite{Patel2024RetrieveTE}. GATher shows better forecasting ability than other approaches and baselines, increasing precision for prioritising the top 200 clinical trial targets to 14.1\% by 2024, an absolute improvement of 3.5\% over PWAS \cite{NarganesCarln2023APA} and 4.3\% over the Open Targets global association score \cite{opentargets2023}.

In summary, GATher, with its GATv3 layer and GATv3HeteroConv, represents an advancement in biomedical graph modelling, leveraging context-aware attention and incorporating clinical trial outcome prediction, offering a robust method to learn from a biomedical graph and rank therapeutic targets within an interpretable framework.

%% file: 02_workflow.tex
\subsection{The Pipeline} 
\label{subsec:pipeline}

\begin{figure*}[p]
    \centering
    \includegraphics[width=\linewidth]{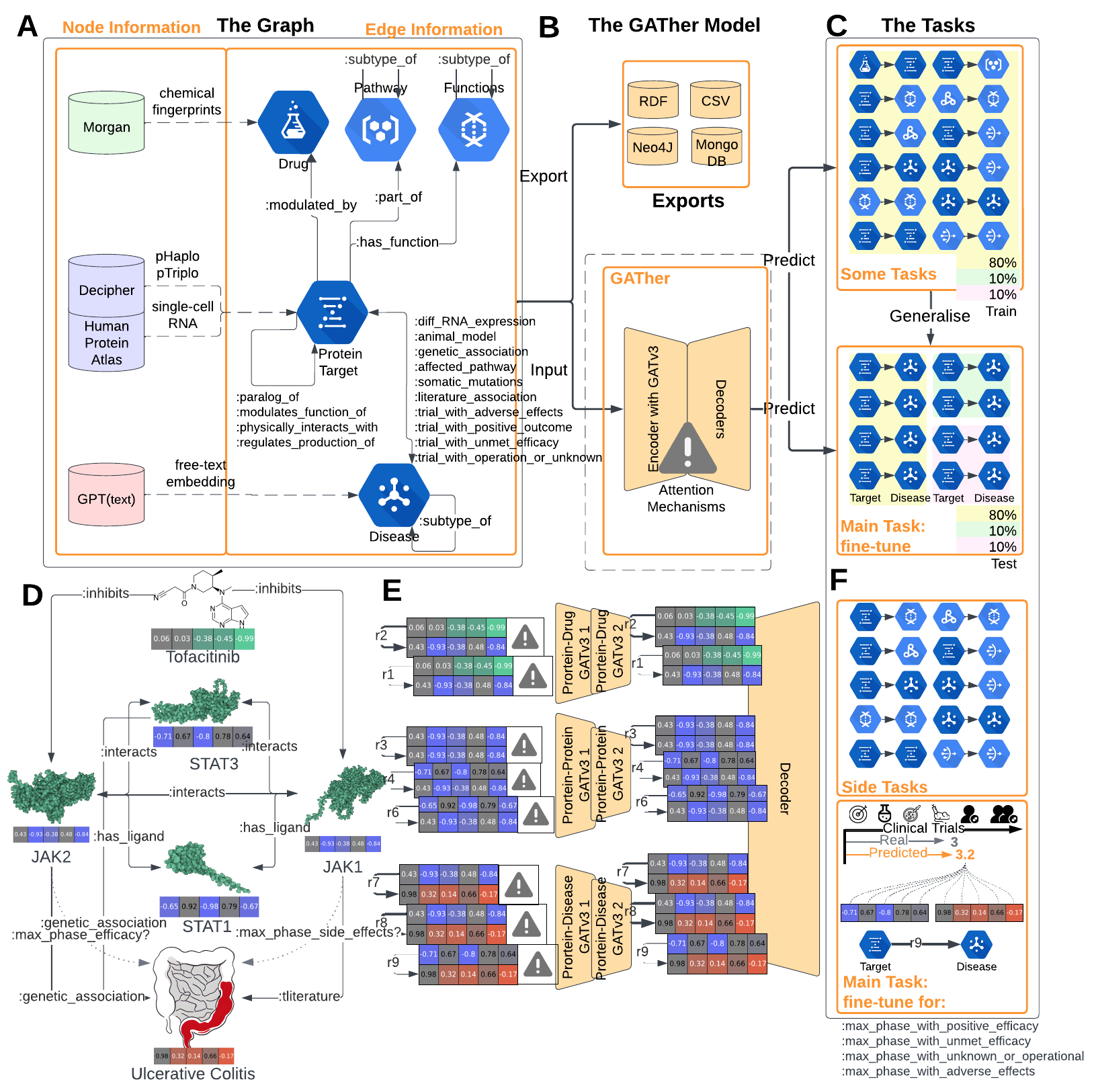}
    \caption{Workflow: The pipeline begins with node and edge input data (\textbf{A} and \textbf{D}), moves through GATher's schematic (\textbf{B} and \textbf{E}), and concludes with model training (\textbf{C)} and \textbf{F}). \textbf{A} displays connections among drugs (green), gene targets (blue), and diseases (red), with edges representing biological and chemical relationships. \textbf{B)} presents the GATher model's schematic, highlighting the encoder-decoder architecture for classifying and regressing edges. \textbf{C)} outlines the training process: training on 80\% of the 2018 data, testing on 10\%, and validating on another 10\% from 2018, with additional validation on a prospective future dataset. The multi-stage training includes initial pre-training and fine-tuning for clinical trial regression. \textbf{D)} details a subgraph illustrating the inhibition of JAK1 and JAK2 by tofacitinib for ulcerative colitis treatment. Entities are shown as coloured vectors with gradient colours: drugs (blue), diseases (red), and drugs (green). \textbf{E)} demonstrates edge formation by transforming node pairs using GATv3’s attention mechanism, with each edge type having a dedicated layer. Nodes integrate features and graph information post-encoding. \textbf{F)} describes multi-task training, where edges are decoded into classification probabilities and regression estimates. The final step is fine-tuning for clinical trial prediction of protein (target)-disease links for the four clinical trial outcomes.
}
    \label{fig:workflow}
\end{figure*}

This section explains the heterogeneous biomedical graph construction and the training and validation pipeline for GATher, incorporating the novel GATv3 layer (Figure \ref{fig:workflow}). The workflow comprises graph construction, node and edge information, schema (Figure \ref{fig:workflow}\textbf{A}), and a subgraph view (Figure \ref{fig:workflow}\textbf{D}), the encoder-decoder architecture (Figures \ref{fig:workflow}\textbf{B} and \textbf{E}), pretraining (Figure \ref{fig:workflow}\textbf{C}), and fine-tuning (Figure \ref{fig:workflow}\textbf{F}). The context-aware attention in GATv3 is shown in Figure \ref{fig:workflow}\textbf{E} and detailed in Section \ref{subsubsec:gatv3}.

In Figure \ref{fig:workflow}\textbf{A}, nodes represent drugs, diseases, proteins, gene functions, and pathways, enriched with node features from databases (Section \ref{subsubsec:engineered}). Morgan fingerprints encode small molecule drug information. Protein targets are characterised by Human Protein Atlas (HPA) \cite{Sjstedt2020AnAO} expression profiles and Decipher \cite{Collins2022, Karczewski2020} genetic features. Disease nodes use Generative pre-trained transformer (GPT) text embeddings (Methods \ref{subsec:graph_data}).

Edges capture protein-protein interactions, pharmacological effects, clinical phase information \cite{Pharmaprojects2023}, and gene-disease associations \cite{SerranoNjera2021TrendyGenesAC, opentargets2023} (Methods \ref{subsec:graph_data}). Key edges are therapeutic edges categorised by clinical trial outcomes: positive efficacy, unmet efficacy, unknown or operational effects, and adverse effect terminations \cite{Pharmaprojects2023}. Pathway participation \cite{Kanehisa2023} and biological processes \cite{Aleksander2023TheGO} provide biological processes.

The encoder-decoder model (Figure \ref{fig:workflow}\textbf{B}) incorporates GATv3 (Section \ref{subsubsec:gatv3}), optimised using a hyperparameter grid search (Methods \ref{sec:methods}, Table \ref{tab:hyperparameters}). The encoder uses a GATv3 layer per edge type with the proposed attention mechanism (Section \ref{subsubsec:gatv3}).

Figure \ref{fig:workflow}\textbf{C} shows initial training on selected edges to create initial embeddings. The workflow progresses from encoding to predictions, divided into training (80\%), testing (10\%), and validation (10\%) using data until 2018, with additional 2024 test set validation. This split ensures performance evaluation. The model's generalisation is tested using unseen 2024 clinical trial data for the four outcomes, allowing the model to capture key patterns.

For task fine-tuning like predicting clinical trial phases (Figure \ref{fig:workflow}\textbf{F}), the model updates pretraining embeddings guided by optimised hyperparameters (Methods \ref{subsec:training}). Therapeutic edges data are split 80/10/10 at the gene-disease level during training, with splits based on hyperparameter configurations (Table \ref{tab:hyperparameters}). This distribution ensures evaluation on known and unseen trials, assessing predictive accuracy and generalisation for phase regression per protein-disease, with performance metrics influenced by hyperparameter tuning (Table \ref{table:best_params}).

Figure \ref{fig:workflow}\textbf{D} visualises a subgraph showing drug-protein-disease interactions, exemplified by tofacitinib with JAK1/JAK2 in ulcerative colitis. Nodes have vector embeddings and engineered features (Section \ref{subsubsec:engineered}). The model aggregates information two steps away in the subgraph for each protein-disease to estimate the trial stage.

Figure \ref{fig:workflow}\textbf{E} illustrates edge transformation, where node embeddings are concatenated and transformed using the attention mechanism (Section \ref{subsubsec:gatv3}). Each task has a dedicated decoder layer (Figure \ref{fig:workflow}\textbf{F}).

A time-split validation strategy assessed generalisation, using datasets before 2018-01-01 for processing and integration, while evaluating predictions on a 2018+ test set, simulating real-world future prediction.

%TC:ignore

%TC:endignore

%% file: 03_eval.tex
\subsection{Performance of GATv3}

\begin{figure*}[p]
    \centering
    \includegraphics[width=\linewidth]{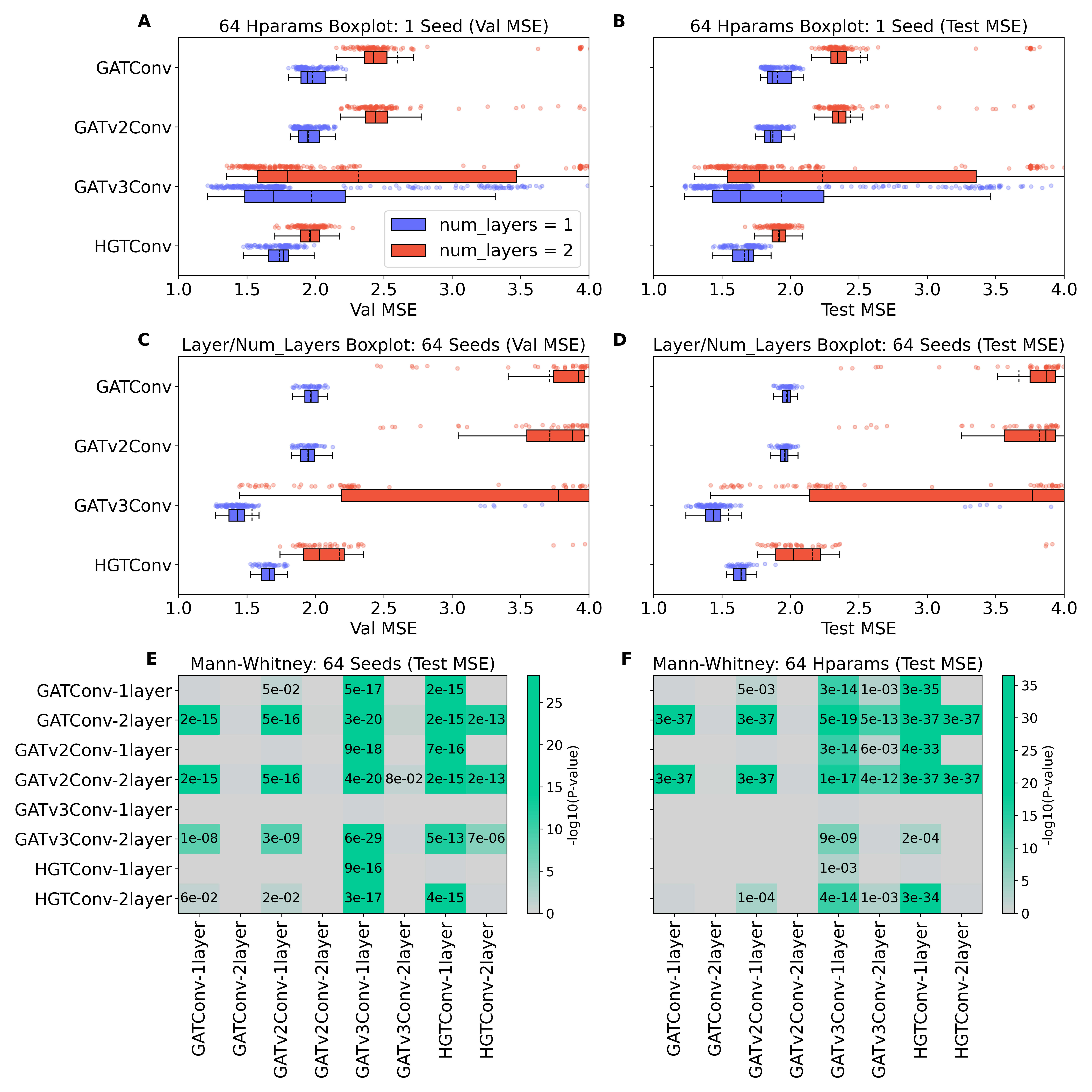}
    \caption{Performance analysis of GATher by layer type and number. (A) Horizontal boxplots showing validation MSE for 1440 hyperparameter sets for each layer type using a single seed. Lower MSE values indicate better performance. (B) Horizontal boxplots showing test MSE for 1440 hyperparameter sets for each layer type using a single seed. (C) Horizontal boxplots showing validation MSE for 1 or 2 layers across 64 seeds, with scatter points for individual MSE values. (D) Horizontal boxplots showing test MSE for 1 or 2 layers across 64 seeds. (E) Heatmap of Mann-Whitney U test results on test MSE from 64 seeds, indicating differences in MSE distributions among layer types and numbers with log-transformed p-values and annotations for significant ones. (F) Heatmap showing Mann-Whitney U test results for test MSE across 1440 hyperparameter sets with one seed, highlighting significant p-values. The figure visualises GATher's performance segmented by layer type and depth, evaluating layers such as GATv3 (ours), GATv2, GAT, and HGT as implemented in Pytorch Geometric \cite{pytorchGeom2019}.}
    \label{fig:eval}
\end{figure*}

GATher, our graph neural network architecture for biomedical data, features GATv3, a graph attention convolution layer designed to capture heterogeneous interactions in biomedical graphs. The GATv3 layer, combined with the GATv3 HeteroConv wrapper, aggregates transformations for each edge type (see Methods \ref{sec:methods}). We conducted two main experiments: Hyperparameter Tuning (Figures \ref{fig:eval}\textbf{A}, \textbf{B}, \textbf{F}) and Stability Analysis (Figures \ref{fig:eval}\textbf{C}, \textbf{D}, \textbf{E}).

In the Hyperparameter Tuning experiment, we compared the GATv3 layer with GAT, GATv2, and HGT layers by testing various parameters, including hidden layer sizes (32, 64, 128), batch normalisation, dropout probabilities (0, 0.1, 0.2), and configurations with one or two layers and one or two attention heads. The GATv3HeteroConv aggregation method was exclusive to the GATv3 layer, resulting in 576 configurations for GATv3 and 288 for each of the other layers, totalling 1440 configurations. We used the sum of the mean squared errors (MSE) for four clinical regression tasks as our primary metric, applying early stopping after ten epochs without improvement. Figures \ref{fig:eval}\textbf{A} and \textbf{B} show validation and test MSE distributions for 64 seeds, illustrating that GATv3 with one layer consistently achieved lower MSE values compared to other configurations.

In Figures \ref{fig:eval}\textbf{A}, \textbf{B}, \textbf{C}, and \textbf{D}, we excluded models with a validation or test MSE greater than 4 to remove poorly performing configurations from the visual analysis; however, these models were retained for the Mann-Whitney U test shown in Figures \ref{fig:eval}\textbf{E} and \textbf{F} to ensure a comprehensive evaluation.

In the Stability Analysis (Figures \ref{fig:eval}\textbf{C}, \textbf{D}), we evaluated the best models from the Hyperparameter Tuning experiment. The best parameters for each layer are shown in Table \ref{table:best_params}. We applied these parameters to all layers to avoid bias, training models using 64 random seeds. Figures \ref{fig:eval}\textbf{C} and \textbf{D} show the variability and performance of each layer type. These boxplots illustrate the spread and consistency of MSE values, reinforcing the superior performance of GATv3 with one layer.

Figures \ref{fig:eval}\textbf{E} and \ref{fig:eval}\textbf{F} present heatmaps of the Mann-Whitney U test for combinatorial comparisons of distributions for different layers and depths (1 and 2 layers). Results show GATv3 had significantly lower MSE than other layers at \( p < 0.05 \). Heatmaps display only significant p-values, with greener shades representing lower p-values.

In Figure \ref{fig:eval}\textbf{E}, the GATv3Conv-1layer configuration shows significant p-values compared to all other layers with one or two depths, with p-values lower than \(9 \times 10^{-16}\). The GATv3Conv-2layer configuration achieves lower MSE values compared to GATConv-2layer, and GATv2Conv-2layer, with fewer significant comparisons. The HGTConv-1layer configuration shows significant p-values against most configurations except itself and GATv3Conv-1layer.

In Figure \ref{fig:eval}\textbf{F}, the GATv3Conv-1layer configuration again shows the most significant p-values, underscoring its robustness for different hyperparameters, with p-values lower than \(1 \times 10^{-3}\) against HGTConv-1layer. The GATv3Conv-2layer configuration shows significant p-values, but less consistently than GATv3Conv-1layer, being significant only against GAT and GATv2 with one or two layers, and HGT with two layers. The HGTConv-1layer configuration outperforms all layers except itself and our GATv3Conv-1layer.

Based on these experiments, shown in Figures \ref{fig:eval} and Table \ref{table:best_params}, GATv3 outperformed all other layers according to the Mann–Whitney U test with a significance threshold of \( p < 0.05 \). HGT with one layer outperformed all layers except itself in Figures \ref{fig:eval}\textbf{E} and \textbf{F}. Single-layer models performed best, suggesting oversmoothing due to high connectivity in biomedical graphs (Appendix, section \ref{subsec:oversmoothing}). These evaluations suggest that graph neural networks designed for biomedical graphs, considering principles of heterophily, can improve performance, as shown by GATv3 with the GATv3 HeteroConv outperforming other layers and HeteroConv wrappers.

%TC:ignore
\begin{table*}[h!]
\centering
\begin{tabular}{l|c|c|c|c}
Layer Name & GAT \cite{GAT2017} & GATv2 \cite{GATv22022} & GATv3 (ours) & HGT \cite{HGT2020} \\
\hline
Val MSE & 1.82 & 1.80 & 1.22 & 1.43 \\
Heads & 2 & 2 & 2 & 2 \\
Heterogeneous Conv Name & HeteroConv & HeteroConv & GATv3HeteroConv & HeteroConv \\
Hidden Channels & 48 & 64 & 128 & 128 \\
Number of Layers & 1 & 1 & 1 & 1 \\
Dropout Probability & 0.2 & 0.2 & 0.1 & 0.2 \\
Use Batch Normalisation & True & True & True & True \\
Use Embeddings & False & False & True & True \\
Use Features & True & True & True & True \\
\end{tabular}
\caption{Best hyperparameters for each of the different layers.}
\label{table:best_params}
\end{table*}

%TC:endignore

\input{03_eval_02_fic}
\input{03_eval_03_heteroconv}

%% file: 03_eval_02_fic.tex
\subsection{First-in-Class and Siren Targets}
To evaluate GATv3's predictive performance within GATher, we assessed its ability to predict clinical phase progression (Phases 1, 2, and 3) for targets that tend to progress positively (first-in-class), increase failures (siren), or remain stuck in the pipeline. First-in-class drugs are the first to modulate a target not previously targeted. Table \ref{tab:first_in_class_drugs} summarises these drugs, diseases, identifiers, gene targets, and FDA approval year from reliable sources.

Siren targets tend to increase trial failures while remaining stuck in the pipeline. A siren target is a gene initially promising but ultimately failing due to repeated efficacy issues. Identified based on historical outcomes, siren targets have a maximum trial phase of 3 or less, with at least two efficacy failures across Phases 2 and 3 for more than one drug and disease, with outcomes like “Terminated, Lack of efficacy” or “Completed, Negative outcome/primary endpoint(s) not met” in PharmaProjects \cite{Pharmaprojects2023}. Once a target is identified as siren, all associated diseases are considered siren target-disease pairs, e.g., CETP and sPLA2. Positively progressing target-disease combinations are annotated in PharmaProjects \cite{Pharmaprojects2023} with “Completed, Early positive outcome” or “Completed, Positive outcome/primary endpoint(s) met” and have increased the maximum trial phase. Post-2018 approved first-in-class targets serve as evaluation examples of positively progressing targets.

\begin{figure*}[p]
    \centering
    \includegraphics[width=\linewidth]{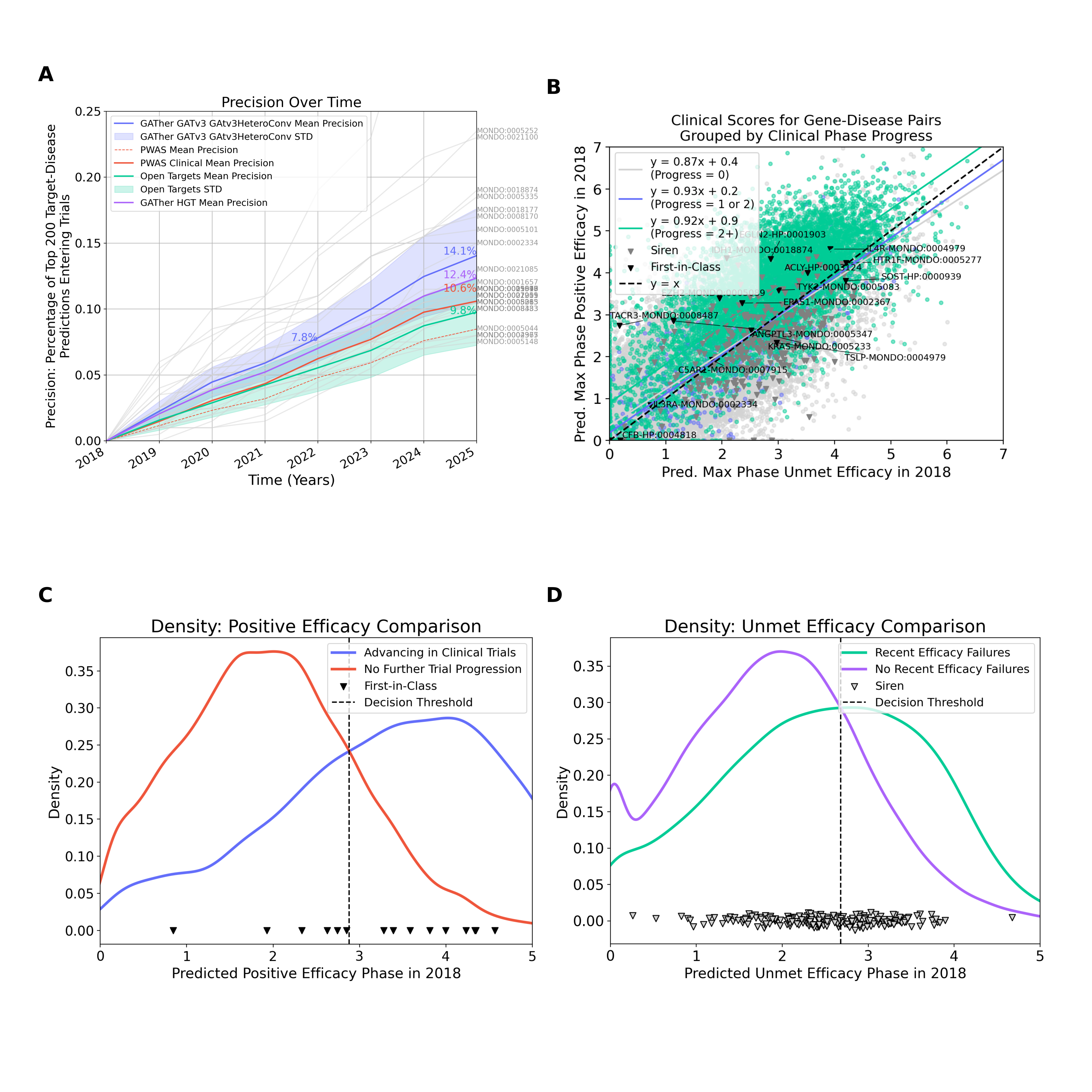}
    \caption{
    Evaluation of clinical phase progression predictions for first-in-class and siren targets.
    (\textbf{A}) Precision over time for predictions of target-disease pairs advancing into clinical trials, showing the GATher median precision and standard deviation across multiple diseases with their individual MONDO terms annotated in grey with grey lines.
    (\textbf{B}) Scatter plot of predicted maximum trial phases with unmet and positive efficacy in 2018, grouped by clinical phase progress. Lines of best fit for each progress category are displayed alongside the `y = x` line. First-in-class targets and siren targets are highlighted with specific symbols.
    (\textbf{C}) Density plot comparing predicted positive efficacy scores between targets advancing in clinical trials and those with no further trial progression. The decision threshold is indicated by the dashed vertical line.
    (\textbf{D}) Density plot comparing predicted unmet efficacy scores between targets with recent efficacy failures and those without. The decision threshold is also indicated from Table \ref{tab:performance}.
    }
    \label{fig:efficacy}
\end{figure*}

Using GATher with GATv3, we predicted the maximum clinical trial phases (with unmet and positive efficacy) for each gene-disease pair, differentiating positively progressing (first-in-class) and failure-increasing (siren) targets. Figure \ref{fig:efficacy} illustrates GATv3's performance in predicting trial stages for first-in-class and siren targets.

GATher's performance in predicting target-disease pairs entering clinical trials demonstrates a marked improvement, with the model achieving a median precision of 7.8\% in 2022 and 14.1\% by 2024 when trained on data up to 2018. On average, 14.1\% of the top 200 predicted targets enter trials, outperforming the best GATher with HGT, which achieved 12.4\% precision. This surpasses other models, including Word2Vec and a multilayer perceptron, as well as Open Targets’ 9.8\% precision from their 2017 release. These results are visualised in Figure \ref{fig:efficacy}\textbf{A}, which shows precision over time for predicted target-disease pairs, with the blue line representing median precision across 20 diseases, and the shaded area indicating standard deviation.

Faster-progressing targets, particularly those predicted to advance by more than two phases, exhibit a positive relationship between positive and unmet efficacy predictions, with slopes of 0.87 for Progress = 0, 0.93 for Progress = 1 or 2, and 0.92 for Progress = 2+. These relationships are visualised in Figure \ref{fig:efficacy}\textbf{B}, where each point represents a gene-disease pair grouped by phase progress. First-in-class targets are represented by black triangles, and siren targets by grey triangles.

Targets with high predicted positive efficacy scores, especially those advancing after 2018 like first-in-class targets, generally have higher scores. The decision threshold separating these groups is close to Phase 2 at 2.32, as detailed in Figure \ref{fig:efficacy}\textbf{C}. This threshold helps distinguish progressing targets from those that remain stuck, enhancing prediction precision to 50\%, recall to 62\%, and MCC to 0.43, with an ROC AUC of 0.79.

Predicted unmet scores for targets with recent failures, such as siren targets, are generally higher, with a decision threshold near Phase 2 at 2.00. Figure \ref{fig:efficacy}\textbf{D} illustrates these findings, showing the model's ability to identify targets likely to fail in trials. The ROC AUC for predicting recent failures is 0.64 using predicted positive scores and 0.65 using predicted unmet estimates.

Table \ref{tab:performance} summarises performance metrics for predicting trial progression, including precision, recall, F1, ROC AUC, and MCC. Decision thresholds are close to Phase 2 due to predicting phase progression per target-disease combination.

To further validate GATher with GATv3, we evaluated Phase 2 to 3 transition prediction performance in Table \ref{tab:performance_phase_2_to_3}. For the Positive Progression group, predicting any phase showed 0.33 precision, 0.62 recall, 0.69 ROC AUC, and 0.24 MCC using predicted unmet scores (Table \ref{tab:performance}). For Phase 2 to 3, precision was 0.31, recall 0.64, and ROC AUC 0.71. Predicted positive scores showed 0.50 precision, 0.62 recall, and 0.79 ROC AUC for any phase. For Phase 2 to 3, precision was 0.47, recall 0.67, and ROC AUC 0.81. In the Unmet Progression group, predicting any phase showed 0.16 precision, 0.58 recall, and 0.65 ROC AUC using predicted unmet scores (Table \ref{tab:performance}). For Phase 2 to 3, precision was 0.15, recall 0.61, and ROC AUC 0.67. Predicted positive scores for any phase had 0.18 precision, 0.48 recall, and 0.64 ROC AUC. For Phase 2 to 3, precision was 0.17, recall 0.51, and ROC AUC 0.66.

These results indicate GATher with GATv3 can distinguish positively progressing targets from Phase 2 to 3 and those increasing failures. The ROC AUCs of 0.65 for unmet failures and 0.79 for positive outcomes for any phase, and 0.67 for unmet failures and 0.81 for positive outcomes for Phase 2 to 3 transitions, support the model's effectiveness and reliability in predicting trial outcomes (Table \ref{tab:performance}) and Phase 2 to 3 transitions (Table \ref{tab:performance_phase_2_to_3}).

%TC:ignore
\begin{table*}[ht]
\centering
\begin{tabularx}{\linewidth}{l|l|l|l|l}
Drug & Disease & Identifier & Gene Target & Year \\
\hline
Iptacopan & Paroxysmal Nocturnal Hemoglobinuria & HP:0004818 & CFB & 2023 \\
Nedosiran & Primary Hyperoxaluria Type 1 & MONDO:0009823 & LDHA & 2023 \\
Fezolinetant & Menopause & MONDO:0008487 & TACR3 & 2023 \\
Deucravacitinib & Plaque Psoriasis & MONDO:0005083 & TYK2 & 2022 \\
Belzutifan & von Hippel-Lindau Disease & MONDO:0002367 & EPAS1 & 2021 \\
Sotorasib & Non-small Cell Lung Cancer & MONDO:0005233 & KRAS & 2021 \\
Evinacumab & Hyperlipoproteinemia Type 2 & MONDO:0005347 & ANGPTL3 & 2021 \\
Tezepelumab & Asthma & MONDO:0004979 & TSLP & 2021 \\
Avacopan & ANCA-associated Vasculitis & MONDO:0007915 & C5AR1 & 2021 \\
Tagraxofusp & Blastic plasmacytoid dendritic neoplasm & MONDO:0002334 & IL3RA & 2021 \\
Daprodustat & Anemia & HP:0001903 & EGLN2 & 2020 \\
Teprotumumab & Graves' Ophthalmopathy & MONDO:0012035 & IGF1R & 2020 \\
Bempedoic Acid & Hypercholesterolemia & HP:0003124 & ACLY & 2020 \\
Lonafarnib & Hutchinson-Gilford Progeria Syndrome & MONDO:0008310 & FNTA & 2020 \\
Tazemetostat & Epithelioid Sarcoma & MONDO:0005089 & EZH2 & 2020 \\
Romosozumab & Osteoporosis & HP:0000939 & SOST & 2019 \\
Voxelotor & Sickle Cell Disease & MONDO:0011382 & HBB & 2019 \\
Lasmiditan & Migraine & MONDO:0005277 & HTR1F & 2019 \\
Ivosidenib & Acute Myeloid Leukaemia & MONDO:0018874 & IDH1 & 2018 \\
Dupilumab & Asthma & MONDO:0004979 & IL4R & 2017 \\
\end{tabularx}
\caption{First-in-class drugs, disease, disease identifier, gene target, and year approved by the FDA}
\label{tab:first_in_class_drugs}
\end{table*}

\begin{table*}[ht]
    \centering
    \begin{tabularx}{\linewidth}{l|l|r|r|r|r|r|r|r|l}
    Group & Pred. Var. & Thresh. & p-val & Prec. & Rec. & F1 & ROC AUC & MCC & Counts \\
    \hline 
    Positive Prog. & Unmet Efficacy & 2.37 & 0.00 & 0.33 & 0.62 & 0.43 & 0.69 & 0.24 & 24784|6277 \\
    Positive Prog. & Positive Efficacy & 3.12 & 0.00 & 0.50 & 0.62 & 0.55 & 0.79 & 0.43 & 24784|6277 \\
    Positive Prog. & Efficacy Distance & 0.62 & 0.00 & 0.35 & 0.58 & 0.44 & 0.70 & 0.27 & 24784|6277 \\
    Positive Prog. & Efficacy Ratio & 0.83 & 0.00 & 0.36 & 0.68 & 0.47 & 0.73 & 0.31 & 24784|6277 \\
    Unmet Prog. & Unmet Efficacy & 2.37 & 5.64E-173 & 0.16 & 0.58 & 0.25 & 0.65 & 0.14 & 27734|3327 \\
    Unmet Prog. & Positive Efficacy & 2.96 & 1.42E-159 & 0.18 & 0.48 & 0.26 & 0.64 & 0.15 & 27734|3327 \\
    Unmet Prog. & Efficacy Distance & 0.45 & 7.17E-17 & 0.13 & 0.48 & 0.20 & 0.54 & 0.05 & 27734|3327 \\
    Unmet Prog. & Efficacy Ratio & 0.74 & 2.13E-46 & 0.13 & 0.62 & 0.22 & 0.58 & 0.08 & 27734|3327 \\
    \end{tabularx}
    \caption{Performance metrics for the best GATher model predictions of clinical trial progression. The table shows results for different prediction variables (Pred. Var.) and their respective thresholds (Thresh.). Metrics include p-value (p-val), precision (Prec.), recall (Rec.), F1 score (F1), ROC AUC (Receiver Operating Characteristic Area Under the Curve), and Matthews correlation coefficient (MCC). The Counts column indicates the number of positive and negative samples used.}
    \label{tab:performance}
\end{table*}

\begin{table*}[ht]
    \centering
    \begin{tabularx}{\linewidth}{l|l|r|r|r|r|r|r|r|l}
    Group & Pred. Var. & Thresh. & p-val & Prec. & Rec. & F1 & ROC AUC & MCC & Counts \\
    \hline 
    Positive Prog. & Unmet Efficacy & 2.37 & 0.00 & 0.31 & 0.64 & 0.42 & 0.71 & 0.25 & 9930|5753 \\
    Positive Prog. & Positive Efficacy & 3.07 & 0.00 & 0.47 & 0.67 & 0.55 & 0.81 & 0.44 & 9930|5753 \\
    Positive Prog. & Efficacy Distance & 0.57 & 0.00 & 0.33 & 0.63 & 0.43 & 0.72 & 0.28 & 9930|5753 \\
    Positive Prog. & Efficacy Ratio & 0.82 & 0.00 & 0.34 & 0.72 & 0.46 & 0.75 & 0.32 & 9930|5753 \\
    Unmet Prog. & Unmet Efficacy & 2.37 & 6.07E-195 & 0.15 & 0.61 & 0.25 & 0.67 & 0.15 & 9981|3012 \\
    Unmet Prog. & Positive Efficacy & 2.96 & 2.26E-244 & 0.17 & 0.51 & 0.26 & 0.66 & 0.16 & 9981|3012 \\
    Unmet Prog. & Efficacy Distance & 0.45 & 1.44E-57 & 0.11 & 0.48 & 0.19 & 0.54 & 0.05 & 9981|3012 \\
    Unmet Prog. & Efficacy Ratio & 0.74 & 9.66E-104 & 0.12 & 0.63 & 0.20 & 0.58 & 0.08 & 9981|3012 \\
    \end{tabularx}
    \caption{Performance metrics for the best GATher model predictions of clinical trial progression from Phase 2 to Phase 3. The table shows results for different prediction variables (Pred. Var.) and their respective thresholds (Thresh.). Metrics include p-value (p-val), precision (Prec.), recall (Rec.), F1 score (F1), ROC AUC (Receiver Operating Characteristic Area Under the Curve), and Matthews correlation coefficient (MCC). The Counts column indicates the number of positive and negative samples used.}
    \label{tab:performance_phase_2_to_3}
\end{table*}
%TC:endignore

%% file: 03_eval_03_heteroconv.tex
\subsection{Performance of GATv3 HeteroConv}

We evaluate our proposed HeteroConv model's performance, attention mechanisms, MSE distribution across configurations, and the relationship between our attention coefficients (alphas, see Methods \ref{subsubsec:heteroconv}), PyTorch Geometric \cite{pytorchGeom2019} attention coefficients, and edge counts. MSE is the sum of squared errors across all model tasks.

\begin{figure*}[p]
    \centering
    \includegraphics[width=1\linewidth]{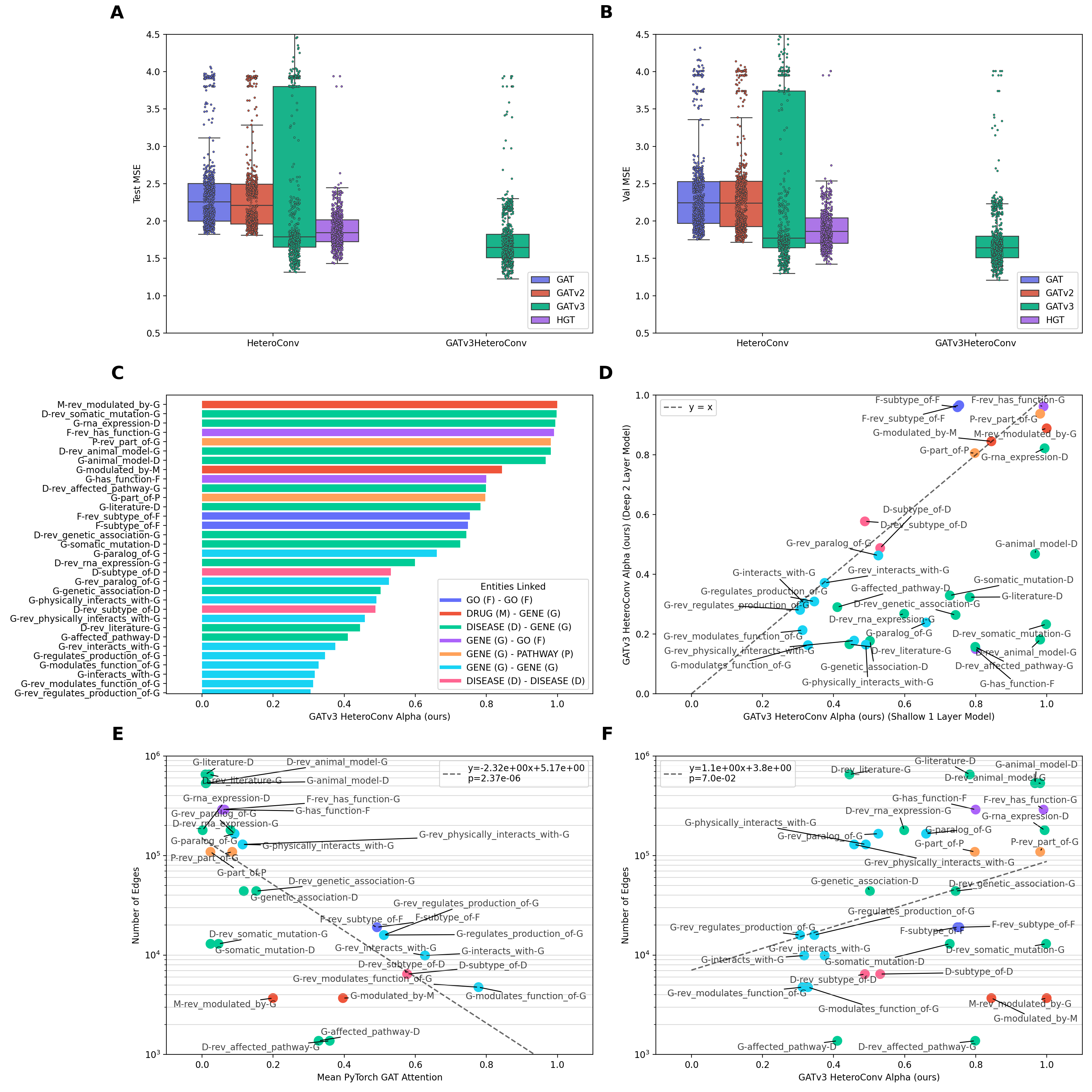}
    \caption{Performance and justification of the GATv3 HeteroConv model. \textbf{A}: Distribution of mean squared error (MSE) values on the test dataset across different attention layers: GAT, GATv2, GATv3, and HGT. This shows the variability and central tendency of MSE for each layer. \textbf{B}: Distribution of MSE values on the validation dataset for different attention layers, indicating consistency of performance across different runs. \textbf{C}: Bar chart of GATv3 HeteroConv Alpha values for different relations. Each bar represents a relation, with its length indicating the alpha value. Colours represent different categories of linked entities as shown in the legend. \textbf{D}: Comparison of GATv3 HeteroConv Alpha values between a shallow 1-layer model and a deep 2-layer model. Each point represents a relation, with its position based on the alpha values in the shallow model (x-axis) and the deep model (y-axis). The dashed line (y = x) shows where the alpha values would be equal for both models. Colours indicate different entity categories as in \textbf{C}. \textbf{E}: Mean PyTorch GAT attention against the number of edges for each relation. The y-axis is the number of edges (log scale), and the x-axis is the mean attention. Colours represent different categories of relations, showing the correlation between attention and edge count. \textbf{F}: GATv3 HeteroConv Alpha values against the number of edges for a shallow 1-layer model. The y-axis is the number of edges (log scale), and the x-axis is the alpha value. The figure shows the correlation between alpha values and edge count for various relations. The colormap for the relations is the same for subplots \textbf{C}, \textbf{D}, \textbf{E}, and \textbf{F}.}
    \label{fig:heteroconv}
\end{figure*}

Figure \ref{fig:heteroconv}\textbf{A} shows test set MSE distribution across attention layers. GATv3 HeteroConv has a lower median MSE than GAT, GATv2, GATv3, and HGT. Our HeteroConv exhibits even lower MSE and reduced variance. Alphas represent relationship importance.

Figure \ref{fig:heteroconv}\textbf{B} depicts validation set MSE. GATv3 HeteroConv again shows lower median MSE. Our HeteroConv further reduces MSE and interquartile range, indicating consistent performance.

Figure \ref{fig:heteroconv}\textbf{C} shows GATv3 HeteroConv Alpha values for entity relations. Bar lengths indicate alphas, colors represent entity categories. High alphas for M-rev modulated by-G, D-rev somatic mutation-G, animal models, and gene-to-function relations highlight significance. Gene-to-gene relations have the lowest alphas. Clinical relations were model outputs.

Figure \ref{fig:heteroconv}\textbf{D} compares shallow 1-layer and deep 2-layer model alphas. Points above y = x have higher alphas in the deep model. Gene-to-function and gene-to-disease associations have higher scores in shallow models.

Figure \ref{fig:heteroconv}\textbf{E} shows Mean PyTorch GAT Attention vs. edge count per relation. Significant negative correlation ($p = 2.37 \times 10^{-6}$) due to softmax causes relations with more edges to have near-zero attention, losing relative importance for critical literature (TrendyGenes \cite{SerranoNjera2021TrendyGenesAC}) and animal model (Open Targets \cite{opentargets2023}) links.

Figure \ref{fig:heteroconv}\textbf{F} shows our GATv3 HeteroConv Alpha values plotted against the edge count for shallow models. Positive correlation ($p = 0.07$) indicates alphas do not suffer softmax normalisation issues. Important literature and animal model relations have higher alphas, reflecting significance across platforms like Open Targets \cite{opentargets2023}, TrendyGenes \cite{SerranoNjera2021TrendyGenesAC}, PWAS \cite{NarganesCarln2023APA}, Pharos \cite{Kelleher2022Pharos2A} and PandaOmics \cite{Kamya2024PandaOmicsAA}.

Overall, our proposed HeteroConv model shows improved performance and stability, with better handling of relationship importance compared to PyTorch GAT's attention mechanisms.

%TC:ignore

%TC:endignore

%% file: 04_feats.tex
\subsection{The Graph Features}
The GATher model integrates engineered features and learned node embeddings to encode graph topological patterns, improving predictive accuracy (Methods \ref{subsec:graph_data}). Gene nodes are characterised using tissue-specific expression from the Human Protein Atlas (approx. 19,200 genes), genetic haploinsufficiency and essentiality annotations, and measures of common essentiality across cell types (Methods).

\begin{figure*}[p]
    \centering
    \includegraphics[width=\linewidth]{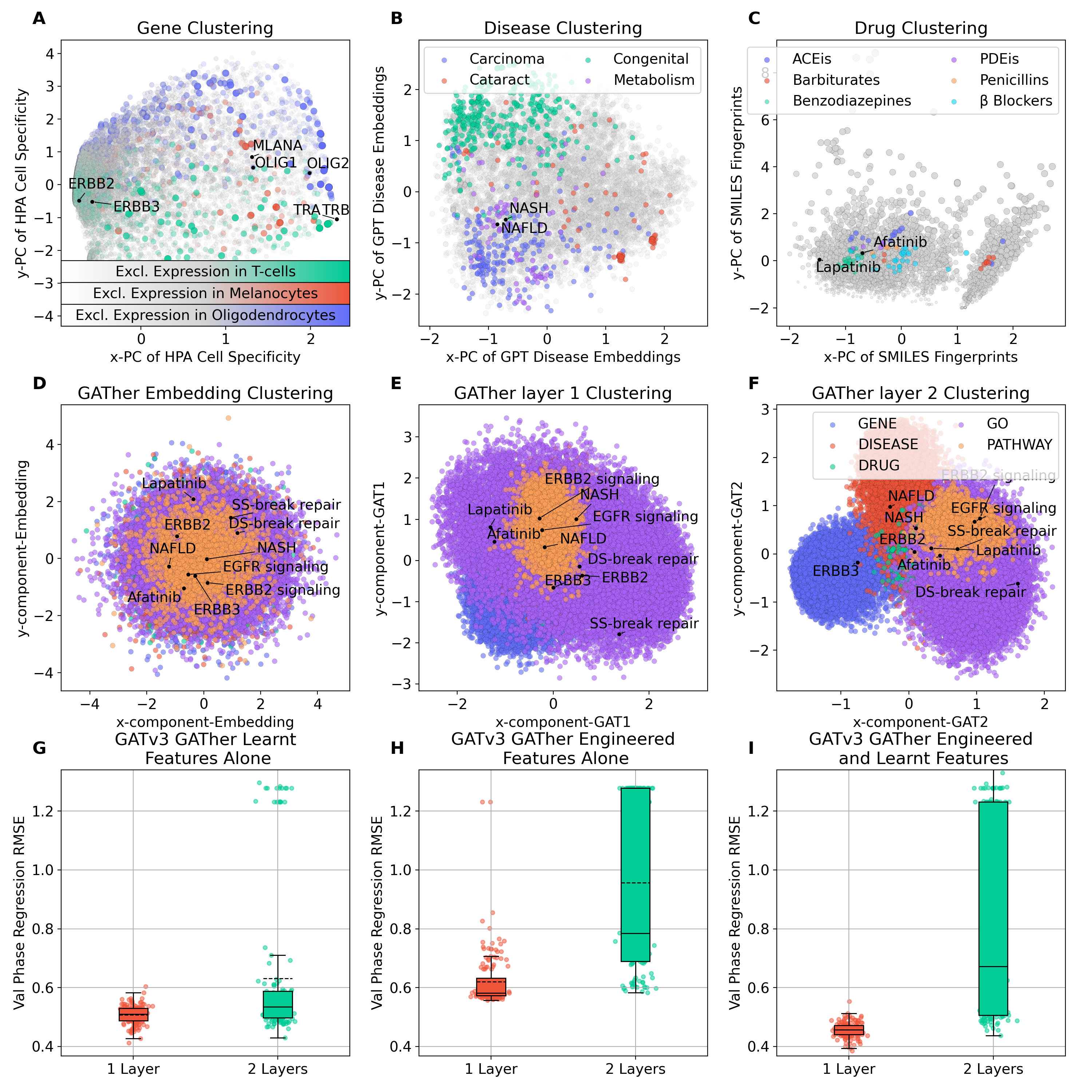}
    \caption{PCA Projections (A to F) and Evaluation of GATher Features (G to I). Engineered Features (A-C): (A) Gene Clustering: PCA of Human Protein Atlas single-cell data shows cell-specific markers in T lymphocytes (blue), melanocytes (green), and oligodendrocytes (red). Highlighted genes include TRA, TRB, MLANA, OLIG1, and OLIG2, with proteins ERBB2 and ERBB3. (B) Disease Clustering: PCA of GPT disease embeddings categorises diseases like carcinoma (blue), congenital disorders (green), and metabolic disorders (purple), marking NAFLD and NASH. (C) Drug Clustering: PCA of SMILES fingerprints identifies drug classes like ACE inhibitors, PDE inhibitors, and beta-blockers, with kinase inhibitors lapatinib and afatinib marked. Learned Features (D-F): (D) GATher Embedding Clustering: PCA shows clusters for genes (blue), diseases (red), functions (purple), pathways (orange), and drugs (green), highlighting processes like EGFR signaling and DNA repair. (E) GATher Layer 1 Clustering: PCA after the first GATv3 encoder. (F) GATher Layer 2 Clustering: PCA after the second GATv3 encoder. Performance Evaluations (G-I, RMSE): (G) Learned Features Alone: Boxplots show RMSE impact by model depth on positive efficacy trials (Score 0.5 preclinical to 4 FDA-approved). (H) Engineered Features Alone: Boxplots show RMSE variation with model depth. (I) Engineered and Learned Features: Boxplots indicate better performance with combined features, balancing manual and automatic extraction.}
    \label{fig:feats}
\end{figure*}

Disease nodes use 256-dimensional Generative Pre-trained Transformer (GPT) language model embeddings. Small molecule drugs are encoded with 1024-bit Morgan fingerprints and learned graph embeddings, initialised randomly and updated during training. Visualisations of learned node embeddings show distinct clusters corresponding to related diseases, pathways, and protein families, identified using Principal Component Analysis (PCA) on engineered features like normalised gene expression profiles (Figure \ref{fig:feats}\textbf{A}), GPT disease embeddings from topological proximity (Figure \ref{fig:feats}\textbf{B}), and chemically diverse molecular fingerprints (Figure \ref{fig:feats}\textbf{C}). Learned graph embeddings projected with PCA show disease, pathway, and protein family groups in the encoder (Figure \ref{fig:feats}\textbf{D}) and after the first (Figure \ref{fig:feats}\textbf{E}) and second (Figure \ref{fig:feats}\textbf{F}) decoder layers. Engineered features (Figures \ref{fig:feats}\textbf{A}-\textbf{C}) and node-learned features (Figure \ref{fig:feats}\textbf{D}) are integrated through concatenation and a feed-forward network, enhancing the encoding layers and improving cluster separation metrics (Figures \ref{fig:feats}\textbf{E}-\textbf{F}).

Quantitative evaluations show combining engineered and learned features in a single-layer model performs best, with a 5th-95th percentile MSE range of 0.409-0.485 on the validation set. Single-layer models using only learned features have a higher error range of 0.472-0.560, while only engineered features perform worst (0.546-0.781). Two-layer models combining both feature types do not improve performance (0.475-1.673). For two layers, only learned features (0.478-1.306) outperform only engineered features (0.582-1.306). Overall, single-layer models integrating engineered and learned representations achieve the narrowest error distribution and lowest percentile values.

The GATher model benefits from complementary use of engineered and learned features. Engineered features provide biological knowledge, ensuring relevance, while learned features capture complex graph structures. The combination accurately estimates drug clinical phases for each gene target-disease pair across four outcomes: positive efficacy, unmet efficacy, adverse effects, and operational/unknown issues, highlighting the effectiveness of combining engineered features with deep learning for target prioritisation and disease recommendation.

%TC:ignore

% https://tableconvert.com/csv-to-latex
\begin{table*}[!ht]
    \centering
    \begin{tabular}{c|c|c|c|c|c|c|c|c|c|c}
    \multicolumn{3}{c}{Model Info} & \multicolumn{4}{c}{Test} & \multicolumn{4}{c}{Validation} \\
    \hline
    Layer & Depth & Feat. & Adverse & Efficacy & Operation & Unmet & Adverse & Efficacy & Operation & Unmet \\
    \hline
        \textbf{GATv3} & 1 & FE & 0.294 & 0.457 & 0.325 & 0.348 & 0.3 & 0.454 & 0.341 & 0.339 \\
        GATv3 & 1 & E & 0.348 & 0.507 & 0.36 & 0.39 & 0.365 & 0.513 & 0.383 & 0.385 \\
        HGT & 1 & FE & 0.324 & 0.522 & 0.366 & 0.406 & 0.36 & 0.511 & 0.397 & 0.381 \\
        HGT & 1 & F & 0.376 & 0.588 & 0.389 & 0.429 & 0.345 & 0.561 & 0.405 & 0.401 \\
        HGT & 1 & E & 0.379 & 0.603 & 0.413 & 0.461 & 0.424 & 0.589 & 0.448 & 0.435 \\
        GATv3 & 1 & F & 0.378 & 0.621 & 0.421 & 0.456 & 0.378 & 0.616 & 0.435 & 0.453 \\
        GATv3 & 2 & E & 0.403 & 0.632 & 0.45 & 0.485 & 0.431 & 0.636 & 0.476 & 0.48 \\
        GATv2 & 1 & F & 0.407 & 0.632 & 0.415 & 0.462 & 0.384 & 0.597 & 0.431 & 0.435 \\
        GATv2 & 1 & FE & 0.418 & 0.644 & 0.431 & 0.487 & 0.396 & 0.616 & 0.457 & 0.461 \\
        GAT & 1 & F & 0.408 & 0.645 & 0.427 & 0.473 & 0.392 & 0.611 & 0.441 & 0.449 \\
        GAT & 1 & FE & 0.418 & 0.655 & 0.436 & 0.494 & 0.402 & 0.625 & 0.461 & 0.468 \\
        GATv2 & 1 & E & 0.445 & 0.676 & 0.447 & 0.512 & 0.426 & 0.649 & 0.479 & 0.487 \\
        HGT & 2 & FE & 0.416 & 0.695 & 0.474 & 0.523 & 0.425 & 0.665 & 0.502 & 0.496 \\
        GAT & 1 & E & 0.459 & 0.697 & 0.463 & 0.53 & 0.436 & 0.667 & 0.491 & 0.498 \\
        HGT & 2 & E & 0.413 & 0.698 & 0.474 & 0.525 & 0.428 & 0.671 & 0.507 & 0.505 \\
        HGT & 2 & F & 0.424 & 0.727 & 0.492 & 0.547 & 0.415 & 0.698 & 0.523 & 0.523 \\
        GATv2 & 2 & F & 0.533 & 0.893 & 0.627 & 0.684 & 0.555 & 0.867 & 0.681 & 0.661 \\
        GAT & 2 & FE & 0.584 & 0.911 & 0.647 & 0.731 & 0.579 & 0.878 & 0.697 & 0.693 \\
        GATv2 & 2 & E & 0.565 & 0.922 & 0.631 & 0.72 & 0.573 & 0.889 & 0.675 & 0.687 \\
        GAT & 2 & F & 0.546 & 0.924 & 0.632 & 0.682 & 0.566 & 0.892 & 0.68 & 0.658 \\
        GATv2 & 2 & FE & 0.551 & 0.927 & 0.633 & 0.715 & 0.568 & 0.895 & 0.69 & 0.7 \\
        GAT & 2 & E & 0.584 & 0.948 & 0.65 & 0.728 & 0.596 & 0.914 & 0.694 & 0.692 \\
        GATv3 & 2 & F & 0.592 & 0.956 & 0.674 & 0.728 & 0.608 & 0.967 & 0.708 & 0.728 \\
    \end{tabular}
\caption{Root Mean Square Error (RMSE) performance across GATher model configurations differentiated by layer depth and type of input data, presented for both test and validation datasets. The RMSE values are categorised by clinical trial outcomes: adverse effects, positive efficacy, operational or unknown issues, and unmet efficacy. This table illustrates the impact of using different model layers and data types—`F` for manually engineered features only, `E` for graph-based embeddings only, and `FE` for a combination of both—on the model's accuracy and predictive reliability across diverse clinical trial phases for target-disease pairs. The models are ranked from top to bottom based on ascending positive efficacy error in the test dataset, highlighting the differential impact of input types on model performance. The best configuration is in bold, the \textbf{GATv3 (ours)}, across all clinical tasks in both test and validation datasets.}
\label{tab:features}
\end{table*}
%TC:endignore

%% file: 05_explain.tex
\subsection{Graph Explanations}
This section explains how GATher predicts therapeutic links between diseases and gene protein-coding targets using integrated gradients \cite{sundararajan2017ig}. Integrated gradients calculate attribution weights for nodes and edges by computing average gradients of the model's output with respect to inputs, scaled by the inputs. These weights indicate each element's importance in the prediction. Nodes represent genes, diseases, pathways, drugs, and gene ontology terms, while edges represent relationships like text mining links, genetic links, drug modulation, pathways, and biological interactions (Section \ref{subsec:graph_data}). Integrated gradients determine each node and edge's contribution by interpolating between a baseline and the actual input, calculating gradients along this path. This identifies influential nodes and edges, providing insights into underlying biological mechanisms, validating predictions, and clarifying identified therapeutic links.

\begin{figure*}[p]
    \centering
    \includegraphics[width=\linewidth]{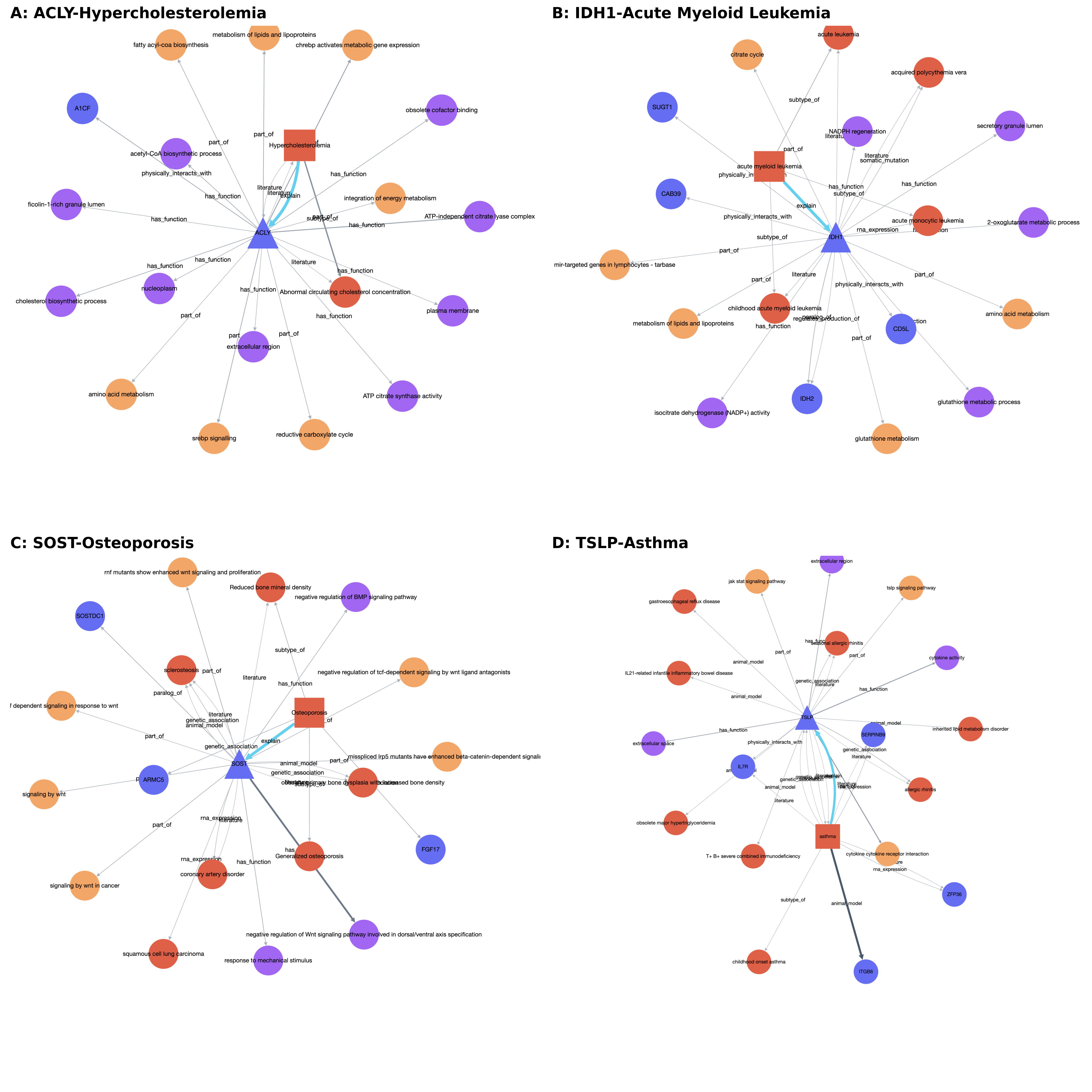}
    \caption{
    Graph attributions for four target-disease pairs: \textbf{(A)} ACLY-Hypercholesterolemia, \textbf{(B)} IDH1-acute myeloid leukaemia, \textbf{(C)} SOST-osteoporosis, and \textbf{(D)} TSLP-asthma. Each network graph illustrates relationships between genes (blue), diseases (red), pathways (orange), and biological processes (purple). Drugs were excluded to focus the model on other entities. Edge widths indicate the significance of the relationships, with thicker edges representing stronger associations based on the attribution score. The top 20 nodes based on total attribution are highlighted, with edge labels specifying the type of relationship between nodes.
    }
    \label{fig:explain}
\end{figure*}

In Figure \ref{fig:explain}, colours denote biological entities: genes are blue, diseases are red, compounds are green, pathways are orange, and biological processes are purple. Targets and diseases are blue triangles and red squares, respectively, with a cyan edge linking the explained gene and disease. Other edges are black, with width indicating integrated gradients attribution weight. Drugs were excluded to focus on biological entities.

Figure \ref{fig:explain}\textbf{A} shows the relationship between ACLY and hypercholesterolemia. ACLY (blue triangle) is connected to hypercholesterolemia (red square) via literature edges. ACLY's role in cholesterol biosynthesis is highlighted by associations with cholesterol biosynthetic process, acetyl-CoA biosynthetic process, and pathways like energy metabolism integration, SREBP signaling, and ChREBP activation (orange circles). Functional roles like ATP citrate synthase activity and ATP-independent citrate lyase complex (purple circles) further illustrate ACLY's contribution. High attribution weights on edges connecting ACLY to hypercholesterolemia and related pathways demonstrate GATher's identification of ACLY as a key target.

Figure \ref{fig:explain}\textbf{B} shows the relationship between IDH1 and acute myeloid leukaemia (AML), a connection supported by the approval of IDH1 inhibitors like Ivosidenib for AML treatment. IDH1 (blue triangle) directly connects to AML (red square) through literature-supported edges, indicating its established role in AML. Its involvement in isocitrate dehydrogenase (NADP+) activity and the 2-oxoglutarate metabolic process underscores its role in cancer cell metabolism. Participation in the citrate cycle, glutathione metabolism, and lipid metabolism (orange circles) is also shown, highlighting IDH1's broader impact on cellular functions. Interactions with proteins such as CAB39, CD5L, and SUGT1 provide additional insights into IDH1's functional network. High attribution weights on edges linking IDH1 to AML and related pathways confirm its significance as a therapeutic target. The integration of RNA expression, somatic mutations, and literature data within the model reinforces the IDH1-AML connection across AML subtypes. The model’s findings align with established therapeutic targets, suggesting that IDH1 modulation remains a viable therapeutic strategy.

Figure \ref{fig:explain}\textbf{C} shows the relationship between SOST and osteoporosis. SOST (blue triangle) directly connects to osteoporosis (red square) through literature edges. Its role is highlighted by associations with response to mechanical stimulus, negative regulation of BMP signaling pathway, and the Wnt signaling pathway (orange circles). Interactions with sclerosteosis and reduced bone mineral density (grey circles) emphasise SOST dysfunction's link to bone disorders. High attribution weights on edges linking SOST to osteoporosis and related pathways indicate SOST's potential as a target. Genetic association, RNA expression, and literature reinforce the SOST-bone disorder link like sclerosteosis. The model highlights SOST's relevance, suggesting modulation as a potential strategy.

Figure \ref{fig:explain}\textbf{D} shows the relationship between TSLP and asthma. TSLP (blue triangle) directly connects to asthma (red square) through literature edges. Its role is highlighted by involvement in the TSLP signaling pathway, JAK-STAT signaling pathway, and cytokine-cytokine receptor interaction (orange circles). Cytokine activity, extracellular region, and space (purple circles) emphasize TSLP's immune response role. High attribution weights on edges linking TSLP to asthma and related pathways suggest TSLP's potential as a target. Genetic association, RNA expression, and literature further support the TSLP-asthma association, including subtypes and comorbidities. The model highlights TSLP's functional role, suggesting modulation as a potential strategy.

The attribution weights align with known biological processes, demonstrating GATher's ability to predict and explain therapeutic gene-disease links. Further research is required to refine subgraph explanations, particularly in identifying key short paths of length 3 or 4 linking entities as this is a limitation of the Integrated gradients algorithm.

%% file: 06_discussion.tex
\subsection{GATher and GATv3}
GATher surpasses models like GAT \cite{GAT2017}, GATv2 \cite{GATv22022}, and HGT \cite{HGT2020}, primarily due to the GATv3 layer. This layer uses context-aware attention, replacing GATv2's LeakyReLU activation with dot product-based scoring, aligning with heterophily principles where shared contexts are more significant than direct links \cite{Kovacs2019Network, Eyuboglu2022MutualIP}. GATv3 incorporates a Transformer-inspired scaling factor \cite{vaswani2017}, stabilising attention score variance and potentially reducing mean squared error (Figure \ref{fig:eval}). Additionally, it includes a bias term and offers weight sharing across node feature transformations, enhancing pattern recognition in biomedical data.

GATher employs an encoder-decoder structure with 1 or 2 layers of graph attention mechanisms, enabling it to work with various architectures like GAT, GATv2, HGT, and GATv3 (detailed in \ref{subsubsec:gatv3}). GATher integrates both engineered and learned features to enhance predictions and interpret biomedical data (Figures \ref{fig:workflow} and \ref{fig:feats}). This combination leverages domain-specific features alongside data-driven graph learning, improving target discovery.

The new features in GATv3, such as context-aware attention, scaling factors, bias terms, and weight sharing, contribute to GATher's improved performance in handling complex biomedical data.

\subsection{The Training Pipeline}

The GATher model (Section \ref{sec:methods}) incorporates diverse node and edge data (Section \ref{subsec:graph_data}), combined with engineered features, to enhance drug target prioritisation. Node data includes single-cell RNA expression for genes, haploinsufficiency and triploinsufficiency insights, disease embeddings from MONDO and HPO ontologies, and molecular structures represented by Morgan fingerprints. Learned node features from graph-based machine learning algorithms offer a comprehensive representation of biological entities and their interactions. Edge data, from sources like MONDO \cite{Vasilevsky2022MondoUD}, HPO \cite{Gargano2023TheHP}, Gene Ontology \cite{Aleksander2023TheGO}, Reactome \cite{reactome2021}, KEGG \cite{Kanehisa2023}, PharmaProjects \cite{Pharmaprojects2023}, and Open Targets \cite{opentargets2023}, include genetic links, protein-protein interactions, and gene-disease associations, resulting in 2,228,495 unique undirected connections. Directed, this amounts to 4,456,990 edges. Figure \ref{fig:feats} shows GATher's effectiveness in predicting and prioritising therapeutic targets using engineered and edge-specific features.

However, reliance on existing databases introduces potential biases due to inaccuracies and variability. Protein-protein interactions often lack sign and directionality \cite{HumanInteractome2020}. Text mining errors \cite{SerranoNjera2021TrendyGenesAC} and incomplete gene ontology associations \cite{Aleksander2023TheGO} highlight the need for careful data integration and refinement.

The training pipeline used the Adam optimiser, learning rate annealing, and L2 weight decay (Section \ref{subsec:training}) to manage large, sparse structures. A full-graph training approach, excluding 40\% of therapeutic edges for validation and testing, was employed to process all edge types in each epoch (Section \ref{subsec:training}), enabling comprehensive attribute integration (Figure \ref{fig:explain}).

We employed hard negative sampling at a ratio of 1 positive to 15 negatives, targeting genes and diseases frequently cited in clinical trials. These nodes, likely to have links, serve as hard negative examples (Section \ref{subsec:training}). This method enhances model specificity by preventing overrepresentation of popular but non-therapeutic targets, ensuring inclusion of druggable, clinically relevant targets, inspired by Huang et al. \cite{huang2023zeroshot}.

GATher's performance was evaluated using the MSE for clinical trial phase regression of target-disease pairs (Section \ref{subsec:evaluation}), visualised in Figure \ref{fig:eval}. The GATher model, incorporating the GATv3 layer, showed variable performance across different experiments, with MSE values shown in Figure \ref{fig:eval} subplots \textbf{A}, \textbf{B}, \textbf{C}, and \textbf{D}. Statistical analysis using the Mann-Whitney U test (Figure \ref{fig:eval}\textbf{E} and \textbf{F}) demonstrated that GATv3 significantly outperformed other layers such as GAT, GATv2, and HGT in both test and validation sets, across various random seeds and hyperparameters. These results demonstrate the GATv3 layer's role in enhancing the performance of the GATher model.

\subsection{The Explanations}
The GATher network, using data up to January 2018, identifies ATP citrate lyase (ACLY) as a key therapeutic target for hypercholesterolemia (Figure \ref{fig:explain}A). ACLY's involvement in cholesterol and acetyl-CoA biosynthesis pathways suggests its role in reducing LDL cholesterol levels. This aligns with Ray et al. \cite{Ray2019SafetyAE}, who detailed the efficacy of bempedoic acid, an ACLY inhibitor approved by the FDA in 2020 (see Table \ref{tab:first_in_class_drugs}).

Similarly, Figure \ref{fig:explain}B depicts the connection between isocitrate dehydrogenase 1 (IDH1) and acute myeloid leukaemia (AML). DiNardo et al. \cite{Dinardo2018DurableRW} showed that ivosidenib, an IDH1 inhibitor, effectively manages IDH1-mutated AML, leading to its FDA approval as a first-in-class medication (see Table \ref{tab:first_in_class_drugs}).

In Figure \ref{fig:explain}C, the GATher network reveals the relationship between SOST (sclerostin) and osteoporosis. SOST's role is highlighted through associations with response to mechanical stimulus, negative regulation of the BMP signaling pathway, and involvement in the Wnt signaling pathway. High attribution weights linking SOST to osteoporosis and related pathways indicate its potential as a therapeutic target. This aligns with Saag et al. \cite{Saag2017RomosozumabOA}, who demonstrated that romosozumab, a sclerostin inhibitor, significantly reduces fracture risk in postmenopausal women with osteoporosis. Published on October 12, 2017, in The New England Journal of Medicine, the study provided evidence supporting romosozumab's FDA approval in 2019 (see Table \ref{tab:first_in_class_drugs}).

In Figure \ref{fig:explain}D, the GATher network with integrated gradients shows the relationship between TSLP (thymic stromal lymphopoietin) and asthma. TSLP (blue triangle) is directly linked to asthma (red square) through literature-derived edges, highlighting its involvement in the TSLP signaling pathway, JAK-STAT signaling pathway, and cytokine-cytokine receptor interaction (orange circles). Gene ontology terms like cytokine activity and extracellular space (purple circles) emphasize TSLP’s role in immune responses. High attribution weights indicate TSLP as a potential therapeutic target. Corren et al. \cite{Corren2017TezepelumabIA} demonstrated that tezepelumab, a TSLP inhibitor, significantly reduces asthma exacerbations, supporting its FDA approval in December 2021 (see Table \ref{tab:first_in_class_drugs}).

A combination of retrieval-augmented generation (RAG) and large language models (LLMs), alongside graph neural networks (GNNs) that predict and highlight relevant subgraphs, may enhance the explanation of new predictions. However, further efforts are required to refine and fully realise this potential.

\subsection{Limitations and Future Work}

The GATher model demonstrates improvements in predicting clinical trial outcomes compared to the previously reported R2E model by BenevolentAI \cite{Patel2024RetrieveTE}. Specifically, GATher achieved a ROC AUC of 0.69 for predicting unmet efficacy failures and 0.79 for positive efficacy outcomes across any phase transition. For Phase 2 to Phase 3 transitions, GATher achieved a ROC AUC of 0.67 for unmet efficacy failures and 0.81 for positive efficacy outcomes. In comparison, the R2E model reported ROC AUCs of 0.545 for genetics-only data and 0.638 for combined data from genetics and literature sources for the 2020 data \cite{Patel2024RetrieveTE}. Similarly, the inClinico model reported a ROC AUC of 0.88 for Phase 2 to Phase 3 transitions, using a proprietary dataset and model, which makes a clear comparison difficult \cite{Aliper2023PredictionOC}. The transparency and reproducibility of GATher's methods, utilising publicly available outcome data from Open Targets \cite{opentargets2023} and Pharmaprojects \cite{Pharmaprojects2023}, as well as describing its full model architecture and including ablation studies, provide a clear benchmark. While these initial results are promising, further efforts are needed to refine and validate GATher to better predict clinical trial outcomes.

Advancements in graph attention networks for biological use cases should consider aligning with biological principles, such as the concept of homophily emphasised in GATv3 \cite{Kovacs2019Network, Eyuboglu2022MutualIP} (Section \ref{subsubsec:gatv3}). While GATv3 incorporates several modifications compared to GAT \cite{GAT2017} and GATv2 \cite{GATv22022}, the cumulative effects of these changes remain underexplored. Models like GATher show potential beyond clinical outcome regression, including tasks like gene function prediction through transfer learning \cite{Zhou2019TheCC}. Although GATher was fine-tuned for therapeutic edges in this study, it could be useful for predicting drug repurposing for rare diseases \cite{HechteltJonker2023IRDiRCDR}, side effects related to drugs or targets \cite{Krix2022MultiGMLMG} and identifying synthetic lethal links between paralog genes \cite{Vazquez2021AreCS}.

The learned relationships in the model were treated as independent from one another. Future research should investigate whether making these relationships interdependent could reduce the number of learnable parameters, improve prediction accuracy, and decrease training times, like Schlichtkrull et al. demonstrated with basis-decomposition methods in relational graph convolutional networks \cite{Schlichtkrull2017ModelingRD}. This could also result in better performance for the two-layer architecture, which in principle should yield higher expressiveness.  Figure \ref{fig:heteroconv}F illustrates that the alpha values for forward and reverse relationships are quite similar (mean difference of 0.15 out of 1.0, x-axis), suggesting that some of these relationships may not require independent parameters.

Improvements to GATher aim to incorporate more diverse datasets, enhancing the model's robustness and predictive capabilities. This study primarily focused on target-disease relationships and did not include additional connections such as drug-disease links, disease-phenotype associations, and directional protein-protein interactions \cite{Chandak2023}. Integrating 'modulated by' edges from bioactive compounds in ChEMBL \cite{Zdrazil2023TheCD} is expected to improve performance by providing more detailed drug-target interactions. The current reliance on binary, unsigned interactions, weighted by confidence scores \cite{BadiaiMompel2023GeneRN}, limits the accurate representation of signalling cascades, which are essential for understanding complex biochemical pathways. As disease ontologies evolve and definitions are refined \cite{Vasilevsky2022MondoUD}, these changes will impact the annotations of protein functions, drug effects, and disease biology. Incorporating patient-level data \cite{Eduati2018PatientspecificLM} could further personalise target discovery, allowing for more tailored therapeutic approaches based on individual genetic profiles. Future work should focus on creating a more interconnected graph \cite{Chandak2023} by integrating these additional data types, potentially improving representation learning and predictive accuracy.

Regarding the explanations, the integrated gradients algorithm \cite{sundararajan2017ig} currently highlights existing connections to and from source and target nodes, as seen in Figure \ref{fig:explain}. Further work is needed to visualise attention weights and identify potential short paths for new predictions \cite{Himmelstein2017}. Combining RAG and LLMs with GNNs that predict and highlight relevant subgraphs could be a powerful approach for explaining new predictions, though more efforts are required. Moreover, integrating textual data into curated databases often lags behind, introducing delays in incorporating new findings. Here, LLMs could bridge the gap by efficiently extracting information from the literature.

%% file: conclusion.tex
\subsection{Conclusion}
GATher, incorporating GATv3 layers, demonstrates significant improvements in predicting biomedical relationships by leveraging context-aware attention mechanisms and dot product-based scoring. The model effectively integrates engineered features, such as gene expression profiles and chemical fingerprints, with learned features from graph embeddings to manage complex, undirected interactions within biomedical graphs.

Validated through MSE evaluations and Mann-Whitney U tests, GATher consistently outperforms previous models across 64 configurations. The results show improved ROC AUC for distinguishing gene-disease combinations likely to fail or succeed, achieving 0.69 for predicting unmet efficacy failures and 0.79 for positive efficacy outcomes. After training on data up to 2018, the precision for prioritizing the top 200 targets entering clinical trials by 2024 increased to 14\%, an absolute improvement of 3\% over Open Targets.

Furthermore, using Captum, GATher identifies key therapeutic targets and pathways, pinpointing first-in-class targets such as ACLY for hypercholesterolemia and IDH1 for acute myeloid leukemia. These findings align with known outcomes, thereby validating the predictive capabilities of the model.

Looking ahead, future work will incorporate diverse datasets, including patient-level data and dynamic interaction networks, to enhance robustness. Additionally, refining explanatory subgraphs will improve interpretability and expand the application scope of the model. However, the performance of GATher relies heavily on the accuracy and completeness of underlying genetic and clinical data, necessitating ongoing updates and validation to mitigate biases and ensure reliability.

%% file: 07_data_nodes.tex
\subsection{Data}
\label{subsec:graph_data}

\subsubsection{Node Information}
\label{subsubsec:engineered}

\textbf{Gene Expression Features for Targets}. The GATher model uses single-cell RNA sequencing data \cite{Sjstedt2020AnAO} to quantify gene abundance across various cell types. The dataset comprises transcript levels for genes in 81 cell types collected from 31 distinct datasets. Each entry includes the gene symbol, cell type, and nTPM (normalised Transcripts Per Million) expression. Normalisation accounts for sequencing depth and gene length, standardising the representation of gene activity.

To quantify gene expression specificity across cell types, we utilise the Jensen-Shannon specificity (JSS) \cite{Lin1991}. JSS values range from 0 to 1, where values near 1 indicate high specificity to a particular cell type, and values near 0 indicate broad expression across multiple cell types.

The JSS is mathematically defined as:

\[
JSS_i = 1 - \sqrt{\frac{1}{2} \sum_{i=1}^{n} \left( p_i \log_2 (p_i) + \frac{p_i + q_i}{2} \log_2 \left( \frac{p_i + q_i}{2} \right) \right)}
\]

In this formula, \( p_i \) represents the actual gene expression distribution across cell types, while \( q_i \) represents a uniform distribution, assuming equal expression across all cell types. The divergence score is calculated using base-2 logarithms, and the JSS is derived by subtracting the square root of half the divergence from 1. This metric helps identify genes with cell type-specific expression, aiding in understanding their biological roles. Figure \ref{fig:feats}\textbf{A} illustrates a Principal Component Analysis (PCA) 2-dimensional representation of these features.

\textbf{Genetic Features for Targets}. To evaluate target potential, we incorporate four genetic features. Haploinsufficiency Score: This score, sourced from Collins et al. \cite{Collins2022}, predicts a gene's functionality with a single copy, indicating its dosage sensitivity. Triploinsufficiency Score: Also from Collins et al. \cite{Collins2022}, this score assesses the impact of additional gene copies, providing insights into gene duplication tolerance. pLOEUF (predicted Loss-of-function Observed/Expected Upper Fraction) Score: Derived from gnomAD sequencing data \cite{Karczewski2020}, this score evaluates tolerance to loss-of-function mutations, reflecting negative evolutionary selection pressure and potential target safety under inhibition. Common Essential Boolean Flag: Based on findings by Pacini et al. \cite{Pacini2021}, this flag identifies genes essential for cell survival as determined by CRISPR screens. High common essentiality is associated with higher risks of clinical trial failures due to safety concerns \cite{Razuvayevskaya2023WhyCT}.

\textbf{Language Model Features for Diseases}. We utilise a Generative Pretrained Transformer (GPT) to semantically encode disease descriptors from both the MONDO \cite{Vasilevsky2022MondoUD} and HPO \cite{Gargano2023TheHP} ontologies. Disease terms, specifically the Preferred Labels from these ontologies, are processed using a GPT-specific tokeniser, converting them into suitable tokens for model input. The GPT model generates 256-dimensional vectors that capture the syntactic and semantic attributes of these terms. By averaging the final hidden state layer for each term, we produce a single embedding vector. These vectors (depicted in Figure \ref{fig:feats}\textbf{B} after PCA 2-component compression) represent the linguistic characteristics of diseases and are integrated into the GATher model encoder, enhancing disease profile understanding. The counts of distinct nodes by type, including Disease Nodes, are summarised in Table \ref{tab:node_types}.

\textbf{Molecular Structure Encoding with Morgan Fingerprints}. RDKit is employed to convert small molecule drugs into their SMILES strings, a textual representation of chemical structures. These SMILES strings are then transformed into Morgan fingerprints, 1024-variable vectors that describe molecular structures within a two-bond radius. For larger molecules, such as biologics and peptides, we use embedding representations instead of Morgan fingerprints due to their complexity. The 1024-variable vectors are compressed into 2 components using PCA for visualisation. Figure \ref{fig:feats}\textbf{C} shows relevant drug classes identified in this low-dimensional representation.

\textbf{Node Feature Integration and Transformation}. The integration and transformation of node features are important for the GATher model's performance. This process involves combining engineered node features (detailed above) with learnable node embeddings, as illustrated in Figures \ref{fig:workflow}\textbf{A} and \ref{fig:feats}\textbf{A} to \ref{fig:feats}\textbf{I}.

Engineered features are concatenated with learned embeddings. For instance, in a GATher model with a hidden state size of 32, the combined feature vector is processed through two linear layers. These interactions allow the features to combine effectively and reduce their dimensionality to the hidden state size of 32. Specifically, gene features such as HPA data, haploinsufficiency, and triploinsufficiency are combined with their respective node embeddings and compressed to a size of 32. Similarly, drug features represented by Morgan fingerprints and disease features represented by GPT embeddings (initially 256-dimensional) are reduced to a hidden state size of 32, as shown after PCA projection in Figures \ref{fig:feats}\textbf{E} after the first GATv3 layer and \ref{fig:feats}\textbf{E} after the full encoder and second GATv3 layer. This process combines structural information from the graph with biological information from engineered features.

All node embeddings are initialised with a normal distribution. To promote stable learning, the embeddings are regulated by specific parameters: the maximum norm of the embedding vectors is limited to 2 using the L2 norm, preventing any vector from becoming excessively large and ensuring training stability. Additionally, gradients are scaled by the inverse frequency of nodes in the edge lists, emphasising less frequent nodes during learning and ensuring balanced training across all nodes.

%% file: 07_data_edges.tex
\subsubsection{Edge Information}
\label{subsubsec:edgedata}

This section details the types of edges structuring the graph in the GATher model, describing interactions among the biological entities used in our graph. The graph includes interactions among 19,200 gene nodes, 28,005 disease nodes, 18,438 gene function nodes, 1,874 drug nodes, and 2,906 pathway nodes (Table \ref{tab:node_types}). Although modelled as directed edges in PyTorch Geometric \cite{pytorchGeom2019} for message passing purposes (discussed in Section \ref{subsec:model_architecture}), all relationships in the graph are fundamentally undirected. The dataset includes 2,228,495 unique undirected edges, which, when modelled as directed edges for computational purposes, total 4,456,990 edges, including reverse relations for message passing in PyTorch Geometric.

\textbf{Gene to Disease Associations}. The dataset includes associations between 19,200 gene nodes and 28,005 disease nodes, collected from databases such as TrendyGenes \cite{SerranoNjera2021TrendyGenesAC}, Open Targets \cite{opentargets2023}, and PharmaProjects \cite{Pharmaprojects2023} (Table \ref{tab:node_types}). This includes 651,931 text-mined associations from TrendyGenes, which use co-citation networks and machine learning algorithms to identify gene-disease co-occurrences in scientific literature. The methodology of TrendyGenes is detailed in the TrendyGenes publication \cite{SerranoNjera2021TrendyGenesAC}. Open Targets \cite{opentargets2023} provides genetic associations from the Open Targets Genetics Portal, RNA expression data from the Expression Atlas, somatic mutation data from the Cancer Gene Census, and animal model associations from IMPC. Additional sources include Orphanet, UniProt, EVA, SLAPenrich, CRISPR screens, Genomics England, Gene2Phenotype, and ClinGen. We used the association data from December 2017 from the FTP server. For the comparison with our models in Figure \ref{fig:efficacy}, we used the Global Score from Open Targets from this version. From PharmaProjects \cite{Pharmaprojects2023}, the dataset incorporates clinical trial outcome data with associations such as termination due to adverse effects, positive efficacy, unknown or operational data, and unmet efficacy. This data contains 31,004 clinical therapeutic edges. All data utilised in this study was sourced from versions available before 01-01-2018.

The clinical trial data is categorised based on outcomes and phases using specific queries filtering for gene-disease associations with designated trial outcomes. Clinical outcomes are classified into four categories: positive efficacy, unmet efficacy, adverse effects, and unknown or operational, based on trial reports and development statuses. PharmaProjects \cite{Pharmaprojects2023} was chosen over ChEMBL due to its detailed historical records of clinical trial outcomes. Information on node types and edge types, including counts and sources, is presented in Tables \ref{tab:node_types} and \ref{tab:edge_types}. Confidence scores for gene to disease associations are derived from Open Targets, which employs a scoring mechanism based on the strength of the evidence linking genes to diseases \cite{opentargets2023}. Additionally, the confidence score from TrendyGenes \cite{SerranoNjera2021TrendyGenesAC} is calculated using the normalised point-wise mutual information score, which measures the probability of co-occurrence of a gene and a disease relative to the probabilities of the entities occurring independently. Each edge representing clinical trial data from PharmaProjects \cite{Pharmaprojects2023} is assigned an integer representing the maximum phase reached for each gene-disease combination, which serves as both the confidence score and the edge weight. These are categorised into four outcome classes: positive efficacy, unmet efficacy, adverse effects, and unknown or operational, based on specific trial reports and development statuses.

\textbf{Gene to Gene Ontology (GO) Associations}. This dataset includes 289,146 associations between 19,200 gene nodes and 18,438 GO term nodes, as recorded in the GO database \cite{Aleksander2023TheGO}. The GO database categorises biological processes, cellular functions, and molecular locations, providing a structured approach to understanding gene functions. Additionally, there are 19,340 function-function associations within the GO terms, detailing hierarchical and subtype connections, enriching the dataset with a structured view of gene functionalities.

\textbf{Gene to Pathway Associations}. This dataset contains 108,710 associations between 19,200 gene nodes and 2,906 pathway nodes (Tables \ref{tab:edge_types} and \ref{tab:node_types}), as documented in Reactome \cite{reactome2021} and KEGG \cite{Kanehisa2023}. These associations describe the involvement of genes in specific biological pathways and biochemical processes. Additionally, the dataset includes 2,571 pathway-pathway associations, describing interactions and subtype relationships between different pathways, describing the hierarchical and functional relationships between pathways.

\textbf{Gene to Gene Associations}. This dataset comprises various types of interactions among 19,200 gene nodes, including 9,700 protein-protein interactions (PPIs) and 128,122 physical interactions, both catalogued through Pathway Commons \cite{HumanInteractome2020}. Additionally, it includes 165,449 paralogous gene associations identified using the Ensembl database via the \texttt{biomaRt} R package \cite{Rainer2019ensembldbAR}, where default thresholds for homology ensure the accuracy of gene duplication records. The dataset also records 15,890 interactions that regulate gene production and 4,750 interactions that modulate gene function, further documented in Pathway Commons \cite{HumanInteractome2020}. These diverse interactions provide information about the functional and regulatory networks within the human interactome.

\textbf{Disease to Disease Associations}. Our biomedical graph incorporates 7,482 disease-to-disease and phenotype-to-disease associations across 28,005 disease nodes (Table \ref{tab:node_types}), derived from both the MONDO \cite{Vasilevsky2022MondoUD} and HPO \cite{Gargano2023TheHP} ontologies. This dataset includes hierarchical relationships, such as parent-child linkages, and alternative disease categories, mapping out disease taxonomy.. The associations were extracted from the MONDO and HPO databases through BioPortal, providing a depiction of how diseases and phenotypes are interconnected within these ontologies.

\textbf{Drug to Gene Modulation}. The dataset includes 3,693 drug-gene modulation interactions (Table \ref{tab:edge_types}), involving 1,874 drug nodes and their target genes, as reported in PharmaProjects \cite{Pharmaprojects2023}. These interactions describe how drugs modulate protein functions, including both activation and inhibition. The dataset includes pharmacologically active compounds, including small molecules and biologics tested in clinical trials. This information adds to the graph the therapeutic mechanisms and functional impacts of these drug interactions on proteins.

All data processing and integration steps were completed using versions of datasets available prior to 2018-01-01.

\begin{table}
\centering
\caption{Counts of Distinct Nodes by Type}
\begin{tabular}{c|c}
Node Type & Count \\
\hline
Gene Nodes & 19,200 \\
Disease Nodes & 28,005 \\
GO Term Nodes & 18,438 \\
Pathway Nodes & 2,906 \\
Drug Nodes & 1,874 
\label{tab:node_types}
\end{tabular}
\end{table}

\begin{table*}[t]
\centering
\begin{tabularx}{\textwidth}{c|c|c|c}
Edge Type & Database & Entities & Edge Count \\
\hline
Genetic Association & Open Targets Genetics \cite{opentargets2023} & Gene to Disease & 45,377 \\
Somatic Mutation & Cancer Census, IntOGen \cite{opentargets2023} & Gene to Disease & 12,717 \\
Animal Models & IMPC \cite{opentargets2023} & Gene to Disease & 537,007 \\
Text Mining & TrendyGenes \cite{SerranoNjera2021TrendyGenesAC} & Gene to Disease & 655,168 \\
Drug-Gene Interaction & PharmaProjects \cite{Pharmaprojects2023} & Drug to Gene & 3,693 \\
Trial Phase - Adverse Effects & PharmaProjects \cite{Pharmaprojects2023} & Gene to Disease & 2,709 \\
Trial Phase - Positive Efficacy & PharmaProjects \cite{Pharmaprojects2023} & Gene to Disease & 16,142 \\
Trial Phase - Unknown/Operational & PharmaProjects \cite{Pharmaprojects2023} & Gene to Disease & 24,548 \\
Trial Phase - Unmet Efficacy & PharmaProjects \cite{Pharmaprojects2023} & Gene to Disease & 11,381 \\
Diff. RNA Expression & Expression Atlas \cite{opentargets2023} & Gene to Disease & 162,044 \\
Pathway Involvement & Reactome, KEGG \cite{Kanehisa2023} & Gene to Pathway & 108,710 \\
Protein-Protein Interaction & Pathway Commons \cite{HumanInteractome2020} & Gene to Gene & 9,700 \\
Physical Interaction & Pathway Commons \cite{HumanInteractome2020} & Gene to Gene & 128,122 \\
Paralogous Genes & Ensembl \cite{ensembl2023} & Gene to Gene & 165,449 \\
Regulates Production Of & Pathway Commons \cite{HumanInteractome2020} & Gene to Gene & 15,890 \\
Modulates Function Of & Pathway Commons \cite{HumanInteractome2020} & Gene to Gene & 4,750 \\
Gene Function & GO Database \cite{Aleksander2023TheGO} & Gene to Function & 289,146 \\
Function Subtype & GO Database \cite{Aleksander2023TheGO} & Function to Function & 19,340 \\
Pathway Participation & Reactome, KEGG \cite{Kanehisa2023} & Gene to Pathway & 108,710 \\
Disease Subtype & MONDO \cite{Vasilevsky2022MondoUD}, HPO \cite{Gargano2023TheHP} & Disease to Disease & 7,482 \\
Pathway Subtype & Reactome, KEGG \cite{Kanehisa2023} & Pathway to Pathway & 2,571 \\
\end{tabularx}
\caption{Edge types with source databases and edge counts as of the training data in 2018.}

\label{tab:edge_types}
\end{table*}

%% file: 08_methods.tex
\subsection{Model Architecture}
\label{subsec:model_architecture}
GATher introduces a new graph attention mechanism, called GATv3, for our heterogeneous biomedical graph, which modifies and extends models like GAT \cite{GAT2017}, GATv2 \cite{GATv22022}, and HGT \cite{HGT2020} to better capture complex biological data interactions. GATv3 is combined with an aggregation layer, called the Hetero Convolution wrapper, to integrate the information using an attention mechanism across edge types. Traditional graph layers may not adequately represent the diverse interactions in biological data due to their uniform transformation across all connections. GATher addresses this by enabling differential treatment of edges, which allows the model to more accurately reflect the diverse interactions within biomedical graphs. Subsequent sections will outline the broader GATher model framework, the architecture of GATv3, and the Hetero Convolution wrapper.

\subsubsection{The GATher Model Architecture}
\label{subsubsec:gather_architecture}

The GATher architecture features an encoder-decoder structure that transforms node features for the prediction tasks outlined in Figure \ref{fig:workflow}C and F. The node features are a combination of engineered features and learned embeddings, processed by the model as described below.

\textbf{Node Feature Integration and Transformation}. 
The integration and transformation of node features are important components of the GATher model. This process involves combining engineered node features (detailed in Section \ref{subsubsec:engineered}) with learnable node embeddings, as illustrated in Figures \ref{fig:workflow}\textbf{A} and \ref{fig:feats}\textbf{A} to \textbf{I}.

Engineered features are concatenated with learned embeddings. For example, in a GATher model with a hidden state size of 32, the combined feature vector is processed through two linear layers. This transformation facilitates interactions among the features and reduces their dimensionality to match the hidden state size of 32. Specifically, gene features such as HPA data, haploinsufficiency, and triplosufficiency (detailed in Section \ref{subsubsec:engineered}) are combined with their respective node embeddings and compressed to a size of 32. Similarly, drug features represented by Morgan fingerprints and disease features represented by GPT embeddings (initially of size 256) are reduced to a hidden state size of 32 (as shown after a PCA projection in Figures \ref{fig:feats}\textbf{E} after the first GATv3 layer and \textbf{E} after the full encoder and the second GATv3 layer). This process integrates both the structural information from the graph and the biological information from the engineered features, which are normalised as described in Section \ref{subsubsec:engineered}.

All node embeddings are initialised with a normal distribution. To maintain training stability, the embeddings are adjusted with specific parameters: the maximum norm of the embedding vectors is limited to 2 using the L2 norm, preventing any vector from becoming excessively large. Furthermore, we scale gradients by the inverse frequency of nodes in the edge lists, which gives more emphasis to less frequent nodes during learning. This approach ensures that the model learns from both rare and common nodes.

\textbf{Encoder: Sequential Feature Aggregation and Update}.
The encoder transforms node representations through one or two layers of graph attention mechanisms. The encoder layers we tested are GAT \cite{GAT2017}, GATv2 \cite{GATv22022}, HGT \cite{HGT2020}, and our GATv3 (described in Section \ref{subsubsec:gatv3}). They update node features by aggregating information from neighbouring nodes, a process involving the computation of attention scores to create a weighted sum of neighbours' features. This iterative process refines feature representations over successive layers. The model supports multiple attention layers, attention heads, and the option for weight-sharing across different relations, allowing it to navigate between complexity and generalisation. Techniques like batch normalisation, L2 normalisation, and dropout are employed to mitigate over-fitting and maintain training stability. Furthermore, the encoder can selectively ignore certain edge types during training, enabling a routine of pretraining and fine-tuning with a focus on different edge types.

\textbf{Decoder: Predictive Task Execution}.
The decoder, as outlined in Section \ref{subsec:training}, is tasked with translating node representations from the encoder into predictions for various node and edge types. It aggregates features using methods like concatenation or dot product and applies linear transformations followed by suitable activation functions to convert node representations into task-relevant outputs. For classification tasks, a sigmoid activation function is used, whereas for regression tasks, the ReLU activation function is employed.

Overall, GATher combines graph attention mechanisms with task-specific decoders, enabling it to process various node features, both engineered and learnable, for improved predictive capabilities. Section \ref{subsubsec:gatv3} provides detailed insight into the GATv3 convolution layer, and Section \ref{subsubsec:heteroconv} describes the aggregation of the GATv3's outputs for different edge types through a cross-relationship attention mechanism.

\subsubsection{The GATv3 Convolution Layer}
\label{subsubsec:gatv3}

The Graph Attention Network version 3 (GATv3) introduces several advancements over its predecessor GATv2 \cite{GATv22022}. Below, we mathematically describe the main differences between these two versions, with a particular focus on the convolution layer.

\textbf{Context-Aware Attention without LeakyReLU}.
GATv3 replaces the fixed, non-linear LeakyReLU activation in GATv2 with a context-aware attention mechanism. GATv3 dynamically calculates attention coefficients based on the similarity of node features while considering the type of relationship between them. This process involves dot products between transformed node features, coupled with edge-type-specific scaling and bias-adjustment. This method allows GATv3 to capture the distinct characteristics and significance of various relationships in the graph, improving the model's ability to interpret and represent complex interactions.

For nodes \(i\) and \(j\), with feature vectors \(\mathbf{h}_i^{(k)} \in \mathbb{R}^{d_k}\) and \(\mathbf{h}_j^{(k)} \in \mathbb{R}^{d_k}\) at layer \(k\), and for each edge type \(e\), GATv3 computes the attention coefficient \(\alpha_{i,j}^e\) at layer \(k+1\) as follows:

\begin{equation}
\alpha_{i,j}^e = \frac{\exp\left( \mathbf{W}_{\text{context}}^{e\top} (\mathbf{O}_{s}^e \mathbf{h}_i^{(k)} * \mathbf{O}_{t}^e \mathbf{h}_j^{(k)}) \right)}{\sum_{k \in \mathcal{N}(i) \cup \{ i \}} \exp\left( \mathbf{W}_{\text{context}}^{e\top} (\mathbf{O}_{s}^e \mathbf{h}_i^{(k)} * \mathbf{O}_{t}^e \mathbf{h}_k^{(k)}) \right)}
\end{equation}

Where:
\begin{itemize}
    \item \(\mathbf{W}_{\text{context}}^{e\top} \in \mathbb{R}^{d_{k+1}}\) is the transposed weight matrix from the context attention layer specific to edge type \(e\).
    \item \(\mathbf{O}_{s}^e \in \mathbb{R}^{d_{k+1} \times d_k}\) and \(\mathbf{O}_{t}^e \in \mathbb{R}^{d_{k+1} \times d_k}\) are transformation matrices for the source and target nodes, respectively, tailored to edge type \(e\), projecting the node features into a \(d_{k+1}\)-dimensional space.
    \item \( * \) denotes element-wise multiplication.
    \item \(\mathcal{N}(i)\) represents the neighborhood of node \(i\), inclusive of the node itself.
\end{itemize}

GATv3 updates to dot product node features, embracing the concept of heterophily for biological networks—where nodes connected indirectly via shared contexts and mutual interactors often have more in common than those directly linked, contrasting with social networks that operate on homophily, where direct connections denote similarity \cite{Kovacs2019Network, Eyuboglu2022MutualIP}.

\textbf{Normalisation with a Scaling Factor}.
In GATv3, the attention scores are scaled down by the square root of the dimensionality of the node features (\(d_k\)) before the softmax function to ensure proper normalisation. This technique, inspired by the Transformer architecture \cite{vaswani2017}, stabilises the variance of attention scores while preserving softmax normalisation. The scaled attention is given by:
   \[
   \alpha_{i,j} = \frac{\alpha_{i,j}}{\sqrt{d_k}},
   \]

where \( d_k \) is the size of the node feature vectors.

\textbf{Bias in the Attention Mechanism}.
Unlike GATv2, GATv3 adds a bias term in the attention mechanism's linear transformation. This provides an additional degree of freedom, allowing the network to learn more complex patterns.

\textbf{Option for Weight Sharing}.
GATv3 allows weight sharing across the source and target node feature transformations, reducing parameters:

\[
O x_i \text{ if sharing, else } O_s x_i, O_t x_j
\]

Where \(O\) is the shared matrix, \(O_s\) and \(O_t\) are source and target transformation matrices, and \(x_i\), \(x_j\) are node \(i\) and \(j\) features.

In the GATv3 layer, node representations are iteratively updated across layers. At each layer \(k\), the representation for node \(i\), denoted as \(\mathbf{h}_i^{(k)}\), is updated based on its previous layer's representation and the representations of its neighbours. This is achieved using attention coefficients that weigh the importance of each neighbour's contribution. Specifically, the updated representation at layer \(k+1\) is calculated as:

\[
\mathbf{h}_i^{(k+1)} = \sigma\left(\alpha_{i,i}^e \mathbf{O}_{s}^e \mathbf{h}_i^{(k)} + \sum_{j \in \mathcal{N}(i)} \alpha_{i,j}^e \mathbf{O}_{t}^e \mathbf{h}_j^{(k)}\right)
\]

where:
\begin{itemize}
    \item \(\sigma\) is a ReLU non-linear activation function
    \item \(\mathbf{h}_i^{(k)}\) is the representation of node \(i\) at layer \(k\)
    \item \(\alpha_{i,j}^e\) is the attention coefficient between nodes \(i\) and \(j\) for edge type \(e\), determining the importance of node \(j\)'s features to node \(i\)
    \item  \(\mathbf{O}_{s}^e\) and \(\mathbf{O}_{t}^e\) are the transformation matrices for source and target nodes specific to edge type \(e\), projecting the node features into a higher-level representation
    \item \(\mathcal{N}(i)\) represents the neighbourhood of node \(i\), including the node itself
\end{itemize}

Thus, each layer refines the node representations based on the previous layer's outputs and the local graph structure.

\textbf{Layer and Bias Initialisation}.
GATv3's initialisation uses the Glorot method for layer weights to balance gradient scales and sets biases to zero.

\subsubsection{The Hetero Convolution Wrapper}
\label{subsubsec:heteroconv}
There are several differences between the original Hetero Convolution wrapper from Pytorch Geometric \cite{pytorchGeom2019} and our proposed GATv3 Hetero convolution wrapper, GATv3HeteroConv.

\textbf{Edge-Type Specific Attention Mechanism}.
GATv3's Hetero convolution uses an edge-specific attention mechanism that assigns unique attention coefficients \(\alpha^e\) for each edge type \(e\). This is calculated by summing the absolute values of a dedicated weight matrix \(\mathbf{W}_{\text{context}}^e\), and applying a shifted, zero-capped sigmoid function:

\[
\alpha^e = \frac{1}{1 + \exp\left(-\left(\sum \left| \mathbf{W}_{\text{context}}^e \right|\right) + 1\right)}
\]

Note that for \(\sum \left| \mathbf{W}_{\text{context}}^e \right| \leq 0\), \(\alpha^e\) is set to 0. The initialisation of \(\mathbf{W}_{\text{context}}^e\) parameters close to zero (see the GATv3 section above) implies each edge type will initially contribute minimally to the node embeddings, as the zero-capped sigmoid function will output values near zero. As training progresses, if the parameters of \(\mathbf{W}_{\text{context}}^e\) increase, the corresponding \(\alpha^e\) values will rise towards one, indicating that those edge types have gained importance in determining the node embeddings. Thus, the model dynamically adjusts the significance of each edge type based on training data.

\textbf{Weighted Message Passing and Aggregation in Convolution}.
In the GATv3 layer, node representations are iteratively updated across layers. At each layer \(k\), the representation for node \(i\), denoted as \(\mathbf{h}_i^{(k)} \in \mathbb{R}^{d_k}\), is updated based on its previous layer's representation and the representations of its neighbors. This is achieved using attention coefficients that weigh the importance of each neighbor's contribution. Specifically, the updated representation at layer \(k+1\) is calculated as:

\[
\mathbf{h}_i^{(k+1)} = \sigma\left(\alpha_{i,i}^e \mathbf{O}_{s}^e \mathbf{h}_i^{(k)} + \sum_{j \in \mathcal{N}(i)} \alpha_{i,j}^e \mathbf{O}_{t}^e \mathbf{h}_j^{(k)}\right)
\]

where:
\begin{itemize}
\item \(\sigma\) is a non-linear activation function, typically ReLU or LeakyReLU.
\item \(\mathbf{h}_i^{(k)} \in \mathbb{R}^{d_k}\) is the representation of node \(i\) at layer \(k\).
\item \(\alpha_{i,j}^e\) is the attention coefficient between nodes \(i\) and \(j\) for edge type \(e\), determining the importance of node \(j\)'s features to node \(i\).
\item \(\mathbf{O}_{s}^e \in \mathbb{R}^{d_{k+1} \times d_k}\) and \(\mathbf{O}_{t}^e \in \mathbb{R}^{d_{k+1} \times d_k}\) are the transformation matrices for source and target nodes specific to edge type \(e\), projecting the node features into a higher-level representation.
\item \(\mathcal{N}(i)\) represents the neighborhood of node \(i\), including the node itself.
\end{itemize}

\subsection{Training Methodology}
\label{subsec:training}

\subsubsection{Optimisation Strategy}
The model's parameters are optimised using the Adam optimiser with an initial learning rate of 0.01. Learning rate annealing reduces the rate by 50\% if the validation loss does not improve over four epochs, and L2 weight decay at 1e-6 is employed to ensure stability in training large, sparse graphs.

\subsubsection{Full-Graph Training Approach}
We employ a full-graph training approach, processing all edges in each epoch, as opposed to the mini-batch method. This technique processes all edges in each epoch, ensuring continuous information flow and maximising the use of graph context during parameter updates.

\subsubsection{Training Objectives and Negative Sampling}
During the pretraining phase, our model employs mean squared error (MSE) regression to generate initial embeddings for genes and diseases. This includes predicting genetic associations and literature-based connections. We also use weighted binary cross-entropy for classification tasks involving protein-protein interactions, gene ontology (GO) relations, and drug-gene interactions.

The specific pretraining tasks are:
\begin{itemize}
    \item Regressing gene-disease links based on genetic associations and literature.
    \item Classifying drug-gene interactions (rev\textunderscore modulated\textunderscore by), disease hierarchies (subtype\textunderscore of), gene-gene interactions (physically\textunderscore interacts\textunderscore with and paralog\textunderscore of), GO relations (subtype\textunderscore of and rev\textunderscore has\textunderscore function), and gene-pathway associations (part\textunderscore of).
\end{itemize}

In the fine-tuning phase, we focus on predicting clinical trial outcomes using regression analysis. This phase uses data from PharmaProjects \cite{Pharmaprojects2023} and emphasises the following gene-disease regression tasks:
\begin{itemize}
    \item Maximum trial phase with unknown or operational issues.
    \item Maximum trial phase with positive efficacy.
    \item Maximum trial phase with unmet efficacy.
    \item Maximum trial phase with adverse effects.
\end{itemize}

We employ hard negative sampling at a ratio of 1 positive to 15 negatives, targeting genes and diseases frequently cited in clinical trials but without confirmed therapeutic links \cite{bonner2022implications}. These popular nodes are used as hard negatives, which we believe helps the model learn to distinguish true therapeutic links more effectively.

Negative edges are labeled as zero to differentiate them from positive associations. Positive associations are annotated with the clinical trial phase, ranging from 0.5 for preclinical to 5 for FDA-approved therapies, for the gene-disease combinations. his method aims to increase the specificity of the model's predictions and to address topological imbalances within the graph.

For example, if Gene A is commonly mentioned with Disease X in a non-therapeutic context, it serves as a hard negative for Disease X. Conversely, pairing Gene A with Disease Y, with no known association, serves as a random negative. These structured sampling strategies are intended to help the model accurately identify clinically relevant gene-disease pairs, potentially improving prediction precision and reliability, as suggested by Huang et al. \cite{huang2023zeroshot}.

\subsubsection{Training Curriculum}
The training curriculum has two stages. First, pretraining involves all edge types to understand the biomedical graph's structure and node interactions. This stage sets up the initial node embeddings, which are crucial for the next phase. After pretraining, the model undergoes fine-tuning, focusing on clinical therapeutic edges as the primary end point of interest in this work, following strategies similar to those discussed by Huang et al. \cite{huang2023zeroshot}.

\subsection{Evaluation Methodology}
\label{subsec:evaluation}
\subsubsection{Dataset and Performance Evaluation}
We utilised all edges for pretraining, except for 20\% of the clinical therapeutic edges from PharmaProjects \cite{Pharmaprojects2023}, split evenly between validation (10\% of the edges) and testing (10\% of the edges). The test set was solely for final evaluation post-training. Our model's accuracy on quantitative associations, like clinical therapeutic links, was measured using Mean Squared Error (MSE), and Binary Cross-Entropy was used for classification tasks, such as disease ontology hierarchies.

The evaluation of the model's ability to predict target-disease pairs that will subsequently enter clinical development was done using a precision metric, which was measured as the percentage of the top 200 target-disease pairs that entered clinical trials in the following years (see Figure \ref{fig:efficacy}A).

\subsubsection{Model Optimisation and Validation}
To prevent overfitting during model training, we implemented an early stopping mechanism, halting the process if there was no improvement in validation performance after 10 consecutive epochs. For hyperparameter optimisation, we conducted a grid search, exploring various configurations to identify the optimal settings for model performance and generalisation on the validation dataset.

The hyperparameter grid included different dropout rates (0, 0.1, 0.2), hidden channel sizes (32, 48, 64, 128), and the use of batch normalisation (enabled or disabled). We also varied the number of attention heads (one or two), the number of layers (one or two), and the types of layers (GATv3Conv ours, GATv2Conv \cite{GATv22022}, GATConv \cite{GAT2017}, and HGTConv \cite{HGT2020} from Pytorch \cite{pytorchGeom2019}). Additionally, we assessed the impact of different heterogeneous convolution types (GATv3HeteroConv (ours) and HeteroConv from PyTorch \cite{pytorchGeom2019}) and the inclusion of engineered features and learnable embeddings. Table \ref{tab:hyperparameters} summarises the hyperparameters evaluated during the grid search.

\begin{table*}[ht]
\centering
\caption{Hyperparameter Grid Search for Model Optimisation}
\label{tab:hyperparameters}
\begin{tabularx}{\linewidth}{l|X}
\hline
\textbf{Hyperparameter} & \textbf{Values Tested} \\
\hline
Dropout Rate & 0, 0.1, 0.2 \\
Hidden Channels & 32, 48, 64, 128 \\
Batch Normalisation & Enabled, Disabled \\
Attention Heads & 1, 2 \\
Number of Layers & 1, 2 \\
Layer Types & GATv3Conv, GATv2Conv, GATConv, HGTConv \\
Heterogeneous Convolution Types & GATv3HeteroConv, HeteroConv \\
Use Engineered Features & True, False \\
Use Learnable Embeddings & True, False \\
\hline
\end{tabularx}
\end{table*}

The GATher model that achieved the lowest MSE across all clinical trial edge tasks had a dropout rate of 0.1, 128 hidden channels, one layer, and two attention heads, without batch normalisation, achieving a validation mean squared error (MSE) of 1.11, the sum of the MSE for the 4 clinical trial tasks (See Table \ref{table:best_params} for GATv3).

Our analysis included a detailed examination of the effects of these hyperparameters on model performance. The results of the grid search, summarised in Table \ref{tab:hyperparameters}, highlight the influence of each parameter on the validation performance. Furthermore, Figure \ref{fig:eval}\textbf{A} and \textbf{B} present the distribution of validation and test MSEs across different configurations, illustrating the superiority of the optimal setup.

We ensured the reliability of our findings by varying the initialisation strategies across multiple training runs with different random seeds. This approach confirmed the consistency of model performance, reinforcing the robustness of the identified optimal configuration. Figures \ref{fig:eval}\textbf{C} and \textbf{D} show the stability analysis of the selected hyper-parameters in Table \ref{table:best_params} across different random seeds, showing the variance due to different initialisations.

\subsection{Software and Tools}
We implemented our models using PyTorch and PyTorch Geometric \cite{pytorchGeom2019}, running computations on NVIDIA A10G GPUs with CUDA 12.0. To understand what was driving the model predictions, we used the Captum library to analyse node and edge contributions, as shown in Figure \ref{fig:explain}.

%% file: acknowledge.tex
\section*{Acknowledgements}
We thank Dr. John Overington for his feedback and discussions, particularly on the time-split, phase transition, and the distinction between first-in-class and siren targets. We also appreciate the contributions from the target analysts and software engineers for their suggestions and technical support in developing the GATher model. Finally, we acknowledge the support from our colleagues throughout this research.

\section*{Funding}
No funding was received for this work.

\section*{Conflicts of Interest}
All authors are affiliated with Exscientia, a for-profit company developing AI technologies for drug discovery and development.

%% file: 09_misc.tex
\subsection{Oversmoothing}
\label{subsec:oversmoothing}
Over-smoothing in graph-based models occurs when increasing the number of layers causes node embeddings to become too similar, which reduces the model's ability to distinguish between nodes. This phenomenon can affect model performance, particularly in capturing biological relationships in highly connected graphs, such as those representing interactions between genes and diseases.

\begin{figure*}[p]
    \centering
    \includegraphics[width=\linewidth]{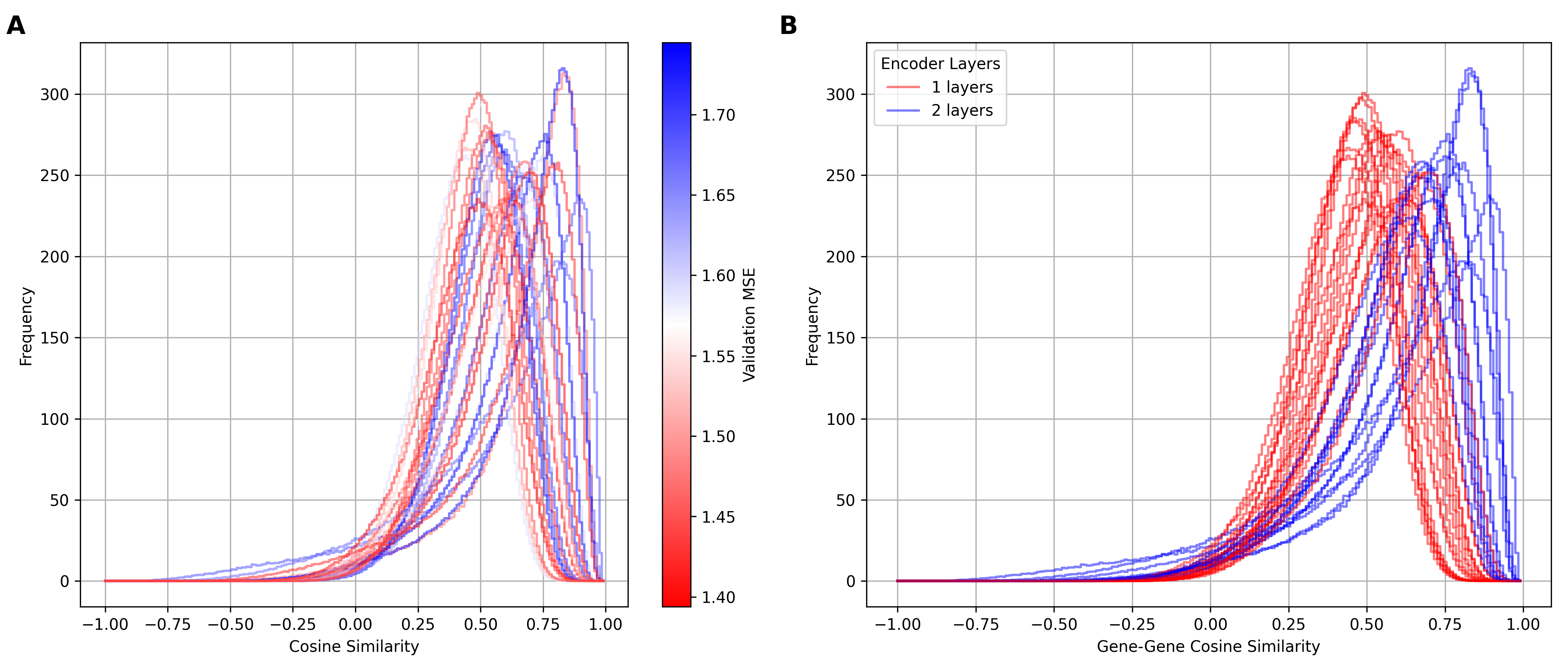}
    \caption{
    Frequency plots of gene-gene cosine similarity in graph-based models, highlighting the issue of oversmoothing. \textbf{A}. The frequency plot shows the distribution of gene-gene cosine similarities across various model configurations, colour-coded by Validation Mean Squared Error (MSE). The gradient from red (low MSE) to blue (high MSE) indicates how specific similarity patterns correlate with predictive accuracy, with over-smoothing potentially leading to higher MSE values. \textbf{B}. The frequency distribution of gene-gene cosine similarities for models with one (red) and two (blue) encoder layers illustrates the impact of model depth in learning different gene embeddings. The comparison shows that increased model depth could lead to oversmoothing, where gene embeddings become overly similar as measured by cosine similarity, thereby diminishing the model's ability to learn meaningful and distinct gene embeddings.
    }
    \label{fig:oversmoothing}
\end{figure*}

Figure \ref{fig:oversmoothing}\textbf{A} shows the frequency plot of pair-wise cosine similarities of gene embeddings across different model configurations, with results colour-coded by Validation Mean Squared Error (MSE). The colour gradient, ranging from red (indicating low MSE) to blue (indicating high MSE), shows the relationship between the distribution of cosine similarities and model performance. High-frequency areas with higher MSE suggest that over-smoothing, indicated by high cosine similarity values, may be associated with increased Validation MSE values, indicating reduced model generalisation on the validation dataset.

Figure \ref{fig:oversmoothing}\textbf{B} presents the frequency distribution of gene-gene cosine similarities for models with one and two encoder layers, depicted in red (1 layer) and blue (2 layers), respectively. This comparison shows how model depth impacts the similarity between gene embeddings. Models with two layers tend to have more pairs of genes with very high cosine similarities, suggesting a potential over-smoothing. This suggests that deeper models are more prone to over-smoothing, which may reduce their ability to learn meaningful gene embedding representations after the encoder.

